\documentclass[a4paper,11pt,twocolumn]{article}

\usepackage[utf8]{inputenc}
\usepackage[T1]{fontenc}
\usepackage[english]{babel}
\usepackage{microtype}
\usepackage{geometry}
\usepackage{parskip}
\usepackage{titlesec}
\usepackage{xcolor}
\usepackage{lineno}

\usepackage{amsmath}
\usepackage{amssymb}
\usepackage{dsfont}

\usepackage{graphicx}
\usepackage{subcaption}
\usepackage{booktabs}
\usepackage{tabularx}
\usepackage{float}
\usepackage{caption}
\usepackage{siunitx}
\usepackage{tikz}
\usetikzlibrary{positioning, fit, arrows.meta, shapes.geometric, backgrounds, calc}

\definecolor{darkblue}{rgb}{0.0, 0.0, 0.55}
\usepackage[colorlinks=true, allcolors=darkblue, urlcolor=darkblue]{hyperref}

\usepackage{csquotes}
\usepackage[numbers, square, sort&compress]{natbib}
\usepackage{authblk}
\usepackage{orcidlink}

\titleformat{\section}{\normalfont\large\bfseries\color{darkblue}}{\thesection}{1em}{}
\titleformat{\subsection}{\normalfont\normalsize\bfseries}{\thesubsection}{1em}{}
\titleformat{\paragraph}[runin]{\normalfont\normalsize\bfseries}{}{0em}{}[.]
\titlespacing*{\section}{0pt}{1.5ex plus 1ex minus .2ex}{1ex plus .2ex}
\titlespacing*{\subsection}{0pt}{1.2ex plus 1ex minus .2ex}{0.8ex plus .2ex}

\title{The Broken Shield of European Palliative Care: Evidence from Synthetic Counterfactuals on Out-of-Pocket Expenditures and Informal Care}
\date{\today}

\begin{document}
	
	\twocolumn[
	\begin{@twocolumnfalse}
		\begin{center}
			\vspace*{-0.5cm}
			{\Large \textbf{The Broken Shield of European Palliative Care: Evidence from Synthetic Counterfactuals on Out-of-Pocket Expenditures and Informal Care}}
			
			\vspace{0.8cm}
			
			{\small 
				\textbf{Pietro Grassi}\textsuperscript{1} \orcidlink{0009-0006-5569-4398},
				\textbf{Edoardo Paperi}\textsuperscript{2} \orcidlink{0009-0002-8948-754X},
				\textbf{Chiara Seghieri}\textsuperscript{3} \orcidlink{0000-0002-3910-7775},
				\textbf{Daniele Vignoli}\textsuperscript{4} \orcidlink{0000-0003-1227-5880}
			}
			
			\vspace{0.3cm}
			
			{\footnotesize 
				\textit{\textsuperscript{1}Honours Student of Data Science at Sant'Anna School of Advanced Studies, Pisa, Italy} \\
				\textit{\textsuperscript{2}Honours Student of Economics at Sant'Anna School of Advanced Studies, Pisa, Italy} \\
				\textit{\textsuperscript{3}Management and Health Laboratory, Institute of Management, Sant'Anna School of Advanced Studies, Pisa, Italy} \\
				\textit{\textsuperscript{4}Department of Statistics, Computer Science, Applications ``G. Parenti'', University of Florence, Italy}\\
			}
			
			\vspace{0.8cm}
			
			\begin{minipage}{0.88\textwidth}
				\small \textbf{Summary}
				
				\vspace{0.2cm}
				
				The transition of end-of-life care to palliative care (PC) sparks intense debate: does it provide economic relief or shift unremunerated labor costs onto families? Evaluating this is hindered by causal inference challenges and skewed healthcare costs. To overcome these limitations, we introduce a Synthetic Data Generation framework. Using pan-European SHARE data (2016--2021), we deploy Tabular Denoising Diffusion Probabilistic Models within a Two-Learner architecture to synthesize high-fidelity digital twins. By including the 2020--2021 lockdowns, we leverage the COVID-19 pandemic to isolate structural inequalities from transient market shocks.
				
				Our findings challenge the strict cost-shifting hypothesis: on average, PC acts as a ``double shield'', truncating out-of-pocket expenditures (liquid toxicity) and informal caregiving shadow values (time poverty). However, quantile treatment models expose a ``broken shield'' for vulnerable households and severe tail events. Non-cancer trajectories drive massive structural penalties that escalate at the distribution's tail, mechanically compounded by physical dependency. Socio-demographics heavily modulate this exposure: lacking a spousal net inflates the burden, rigid gender dynamics trigger labor market ejection, and financial distress acts as a profound multiplier. Institutionally, high-wage Nordic regimes paradoxically impose opportunity costs, while severe penalties in underfunded Eastern systems—mediated by financial distress—drive families toward resource exhaustion. We conclude that while PC is an ethical imperative, its expansion must be decoupled from the oncological paradigm and matched with state-funded long-term care to protect against clinical decline and financial shocks.
				
				\vspace{0.5cm}
				\vspace{0.5cm}
				
				\noindent \textbf{Statements and Declarations:} The authors declare no conflict of interest.
				\vspace{0.5cm}
				
				\noindent \textbf{Funding:}  The authors received no external funding.
				\vspace{0.5cm}
				
			\end{minipage}
		\end{center}
		\vspace{0.8cm}
		\hrule
		\vspace{0.8cm}
	\end{@twocolumnfalse}
	]
	
	
	\section{Introduction}
	
	The economic burden of end-of-life (EoL) care represents one of the most critical challenges for modern healthcare systems and aging societies. As the prevalence of chronic, life-limiting illnesses rises, the prolonged dying process increasingly exposes patients and their families to severe economic shocks \citep{Knaul2018}. Originally conceptualized within oncology as ``financial toxicity'' \citep{Carrera2018}, this drain on household resources is now increasingly recognized as a systemic threat across all prolonged terminal trajectories. Beyond the direct medical costs absorbed by the health system, a substantial ``hidden burden'' remains. This manifests through escalating out-of-pocket (OOP) expenditures and the unremunerated labor demanded of informal caregivers, driving ``time poverty'' and productivity losses for families \citep{Gardiner2014, Urwin2021}.
	
	Palliative care (PC) has emerged as the gold-standard intervention to alleviate suffering at the EoL \citep{Who2020}. While health economics literature demonstrates that PC integration successfully reduces acute healthcare utilization \citep{May2018, Fassbender2009}, its broader societal impact remains debated. A critical question arises: does the shift toward PC genuinely protect families, offering comprehensive relief, or does it merely trigger a cost-shifting phenomenon, where the economic burden is offloaded from the state onto the shoulders of informal caregivers? \citep{GomezBatiste2017, Gardiner2020}. Specifically, we question whether this cost-shifting dynamic escalates into a structural failure for patients requiring continuous, 24/7 supervision, where shifting away from fully-funded acute hospital wards might create an insurmountable care void for the family.
	
	Furthermore, this protective potential is not uniformly distributed, leading to what we define as disease-based, institutional, and socio-demographic inequality \citep{Grassi2026}. Historically designed around predictable oncological trajectories \citep{Murray2005}, PC pathways frequently reach non-cancer patients too late or with inadequate domiciliary support \citep{Radbruch2020, Sleeman2019}. This clinical discrimination intersects with the fragmentation of European welfare regimes and individual household vulnerabilities: we hypothesize that the economic shock of dying at home is profoundly moderated not only by state funding and clinical comorbidities, but by gender dynamics, the availability of a spousal safety net, and highly granular socio-economic determinants such as accumulated private wealth, homeownership, and subjective financial distress.
	
	Methodologically, estimating the true Average Treatment Effect (ATE) of PC on the net economic burden is hindered by the fundamental problem of causal inference \citep{Holland1986} and the unethical nature of withholding PC in randomized trials. Additionally, traditional observational models struggle with the zero-inflated, fat-tailed distributions typical of healthcare costs \citep{ManningMullahy2001, Mihaylova2011}, as well as the complex confounding inherent in terminal trajectories \citep{Fischer2024, AtheyImbens2017}. To overcome these limitations, we bridge health economics and causal machine learning using longitudinal data from the Survey of Health, Ageing and Retirement in Europe (SHARE). By embedding Tabular Denoising Diffusion Probabilistic Models (TabDDPM) \citep{Kotelnikov2023} within a non-parametric T-Learner architecture \citep{Kunzel2019}, we synthesize high-fidelity digital twins to simulate a complete counterfactual manifold, robustly estimating PC's impact on OOP costs and the shadow value of informal care. Furthermore, by encompassing EoL trajectories from 2016 to 2021, our data uniquely spans the COVID-19 pandemic. We leverage this unprecedented exogenous shock to actively isolate deeply rooted structural inequalities from transient pandemic-induced market volatility.
	
	The remainder of this article is structured as follows. Section 2 outlines the theoretical framework. Section 3 details the empirical methodology, the economic monetization strategy, and the causal AI architecture. Section 4 presents the distributional shifts and the causal estimations. Section 5 discusses the findings and their policy implications. Finally, Section 6 concludes the study.
	
	\section{Theoretical Framework}
	
	\subsection{The Economics of EoL Care: Substitution vs. Comprehensive Relief}
	
	In health economics, the reallocation of EoL care from traditional to palliative, especially domestic, settings has traditionally been analyzed through the lens of resource optimization and cost containment \citep{Smith2014}. However, standard economic evaluations frequently fail to account for the economic burden, often treating informal caregiving as a free resource \citep{Hoefman2013, Husereau2022}. On the contrary, our framework adopts a total economic burden perspective that monetizes both OOP expenses and the opportunity costs of informal care (time poverty) \citep{VanDenBerg2004, Knaul2018, Carrera2018}.
	
	The literature presents two competing hypotheses regarding the economic impact of domiciliary PC on households. The cost-shifting hypothesis posits that while PC reduces formal healthcare expenditures, it essentially transfers the caregiving burden to families, substituting paid medical labor with unpaid domestic labor \citep{Gardiner2020, Urwin2021}. Conversely, the comprehensive relief hypothesis builds upon the established health economics consensus that proactive PC successfully prevents acute hospital escalations and reduces formal medical costs \citep{May2018, Cassel2018}. Extending this systemic efficiency to the household level, this hypothesis suggests that well-structured PC provides a ``double shield'': replacing chaotic medical crises and informal family labor with structured professional support, thereby mitigating both financial and temporal shocks simultaneously. Our study explicitly tests these competing macroeconomic hypotheses across the European landscape.
	
	Crucially, we theorize that the validity of these competing paradigms is strictly dependent on the extremity of the patient's dependency, exposing a potential ``catastrophic tail'' exception. European universal health systems historically exhibit a strong institutional bias, where standard acute hospital care fully absorbs the massive costs of inpatient lodging and round-the-clock (24/7) nursing labor \citep{Bekelman2016, French2017}. We hypothesize that at extreme percentiles of clinical severity—where patients require continuous, uninterrupted supervision—episodic domiciliary PC visits or co-pay-reliant residential hospices cannot mathematically substitute the 24-hour care void left by the hospital. Consequently, for the most grueling terminal trajectories, we expect the protective shield of PC to collapse entirely. At this extreme margin, the transition to PC is hypothesized to act as a mechanism of radical cost-shifting, forcing the household to internalize the massive temporal and financial shock of providing or purchasing 24/7 care.
	
	\subsection{Hidden Costs and Socio-Demographic Vulnerabilities}
	
	The economic vulnerability of households is particularly exacerbated in the context of prolonged, non-malignant illness trajectories, such as dementia or extreme frailty. These complex conditions demand staggering levels of daily informal care, exposing caregivers to significantly higher odds of financial hardship compared to cancer trajectories \citep{Gardiner2020}. From a microeconomic perspective, the demand for this essential daily care behaves as perfectly inelastic: households cannot opt out of providing assistance for their dying relatives. 
	
	This inelastic demand interacts critically with the household's socio-demographic structure. Health economics literature demonstrates that severe medical shocks trigger profound intra-household responses, where the presence of a partner fundamentally shapes the family's economic and behavioral adjustment \citep{Fadlon2019}. Extending this logic to the macroeconomic framework that specifically isolates OOP medical risks among single individuals—who must absorb liquid toxicity without the income or labor supply buffering of a partner \citep{DeNardi2016}—we hypothesize that single patients will experience inflated economic penalties. Likely lacking this immediate intra-household safety net, they may be forced to rely heavily on costly formal care markets or distant relatives.
	
	Simultaneously, the volume of informal caregiving hours operates as a compounding penalty on the caregiver's personal economic trajectory, manifesting with profound gender asymmetries \citep{Vignoli2025, Floridi2022}. As demonstrated by \citet{Heger2020}, women tend to balance caregiving by exploiting part-time arrangements, reducing their working hours. While this forced flexibility may artificially soften immediate OOP expenditures, it causes profound ``labor market scars'' that persistently depress their future earning capacity. Conversely, male caregivers—facing rigid ``ideal worker'' cultures and corporate stigma—tend to lack flexible intensive-margin adjustments \citep{Lott2016, Chung2020}.
	
	\subsection{State Underfunding, Acute Financial Distress, and the Wealth Buffer}
	
	The economic shock generated by EoL care needs is profoundly mediated by the institutional architecture of Long-Term Care (LTC) financing. In the absence of adequate universal health coverage, traditional economic theory suggests that households could hedge against the financial risks of EoL dependency by purchasing private LTC insurance \citep{Canta2012}. However, the private LTC insurance market in Europe is heavily undersupplied \citep{CostaFont2015, Albertini2025}. This structural failure is rooted in the foundational theory of uninsurable medical uncertainty \citep{Arrow1963}, specifically exacerbated in the LTC sector by the extreme aggregate uncertainty of forecasting care needs decades into the future, a temporal horizon that precludes efficient actuarial pricing \citep{CostaFont2015}.
	
	This dual institutional failure creates a structural void. Deprived of both functional private risk pooling and adequate public LTC entitlements, households are systematically forced to act as the insurers of last resort. Within this theoretical void, we posit that accumulated private wealth becomes the primary surrogate insurance mechanism, acting as a direct liquid buffer against huge OOP expenditures. Conversely, illiquid structural assets like homeownership cannot be easily liquidated to cover immediate EoL OOP costs, precisely because the primary residence is actively required as a physical setting for care. Instead, homeownership provides a crucial baseline of spatial and economic stability—eliminating ongoing rent burdens and offering the physical infrastructure necessary to sustain caregiving. However, raw wealth metrics often fail to capture acute household liquidity constraints. Therefore, we hypothesize that subjective financial distress—the inability of a family to make ends meet—acts as a massive independent driver of the net burden, effectively neutralizing theoretical average protections. Furthermore, drawing on the economics of extreme health shocks and asset depletion \citep{DeNardi2016, Poterba2017}, we hypothesize a theoretical ceiling to the wealth buffer: at the tail of the distribution, the compounding costs of prolonged dependency exhibit such massive variance that they can overwhelm even the accumulated wealth of affluent families, rendering OOP trajectories wildly unpredictable and universally damaging.
	
	\subsection{Comparative Welfare Regimes and Institutional Inequality}
	
	To contextualize the heterogeneity of PC's protective effects, we draw upon \citet{EspingAndersen1990}’s comparative welfare state theory, extended to encompass variations in health and LTC systems \citep{Bambra2011}.
	
	We classify the European landscape into four distinct regimes. The Continental regime is grounded in Bismarckian social insurance models featuring well-developed, institutionalized LTC infrastructures. The Southern regime is historically defined by fragmented public provision and a deep-rooted structural reliance on the family unit \citep{Ferrera1996}, forcing households to act as ultimate welfare providers. The Eastern regime shares this intense familialistic burden but is compounded by chronic systemic underfunding and widespread economic vulnerability \citep{Rechel2011}. We hypothesize that the severe economic penalties associated with transitioning to home care in Eastern regimes are not merely administrative failures, but are profoundly mediated by the underlying acute financial distress of their populations. For these highly vulnerable households, the compounding costs of care will ultimately lead to absolute resource exhaustion at the extreme tail of the distribution. Finally, the Nordic regime is characterized by universalistic, tax-funded systems with robust formal domiciliary networks. While traditional literature assumes this state defamilialization theoretically maximizes the protective shield of PC, we hypothesize a structural paradox. Comparative analyses highlight a creeping ``de-universalisation'' and ``re-familialisation'' within Nordic eldercare \citep{Szebehely2018}: because Nordic economies are characterized by near-universal dual-earner models and exceptionally high full-time female employment rates compared to the OECD average \citep{Thevenon2011, Korpi2013}, any residual informal caregiving required to bridge the gaps in formal provision translates into a massive monetized opportunity cost. Consequently, we hypothesize that the transition to home-based PC in these high-wage markets will paradoxically impose an economic penalty on families forced to withdraw from the labor market.
	
	\subsection{Clinical Trajectories and Disease-Based Inequality}
	
	The protective capacity of PC is also constrained by the clinical nature of the dying process. The cancer trajectory is typically characterized by a predictable decline following a prolonged period of maintained physical function. Historically, PC models were explicitly designed to accommodate this specific trajectory \citep{Murray2005}. Conversely, patients dying from organ failure or extreme frailty follow pathways characterized by prolonged, unpredictable functional decline and intense dependency \citep{Etkind2017}. This severe physical deterioration acts as a deterministic biological engine that mechanically compounds the economic burden across all settings. Furthermore, this trajectory is often accompanied by multiple baseline clinical comorbidities, which directly inflate OOP outlays. Because EoL systems are structurally misaligned with complex, non-cancer trajectories, we expect their economic penalty to manifest as a massive structural driver across the entire distribution, before violently exploding into severe tail events at the extremes. For the most complex cases of prolonged dwindling, the standard protective capacity of PC will collapse, exposing the structural limits of oncologically-designed care networks.
	
	\subsection{Exogenous Macro-Shocks and the Pandemic Stress Test}
	
	The fundamental architecture of EoL care is heavily reliant on the seamless interaction between formal medical supply and informal domestic networks. Consequently, estimating structural welfare losses requires controlling for severe exogenous macro-shocks capable of displacing these markets. The COVID-19 pandemic represents the most profound contemporary stress test of this dynamic. 
	
	During global lockdowns, formal LTC services, domiciliary nursing, and institutional facilities became largely inaccessible or were perceived as high-risk environments \citep{Werner2020}. Aligning with the epidemiological consensus that the pandemic acted primarily as a magnifier of existing social gradients rather than a novel equalizer \citep{Bambra2020}, we theorize that the pandemic induced a forced ``re-familialisation'' of EoL care across all of Europe. By temporarily fracturing the formal institutional shields of the Continental and Nordic regimes, the shock forced a continent-wide reversion to the familialistic default typical of Southern and Eastern Europe. Furthermore, due to supply chain disruptions and the sheer scarcity of formal care alternatives, we hypothesize that this exogenous shock injected massive price volatility into shadow care markets, inflating the economic burden predominantly at the most severe percentiles of dependency \citep{Grabowski2020}. By mathematically isolating this pandemic fixed effect, our framework aims to disentangle the transient friction of crisis-era healthcare from the permanent, structural failures of European welfare regimes.
	
	\subsection{Causal Inference and the Digital Twin Paradigm}
	
	Evaluating the causal impact of PC on economic outcomes faces huge methodological hurdles. Conducting randomized controlled trials (RCTs) regarding PC is considered unethical \citep{Aoun2005}, while observational data suffer from complex confounding by indication (e.g., wealthier or sicker patients might differentially access PC) \citep{Fischer2024, AtheyImbens2017}. Furthermore, the right-skewness and zero-inflation typical of healthcare cost distributions violate the assumptions of standard linear estimators \citep{Mihaylova2011}.
	
	To address this, we embed our theoretical framework within the paradigm of causal machine learning \citep{Pearl2009, Chernozhukov2018}. We conceptualize the causal effect using the potential outcomes framework \citep{ImbensRubin2015}, where the ATE is defined as $\mathbb E[Y_1 - Y_0]$, with $Y_1$ representing the outcome with PC and $Y_0$ the outcome under standard care.
	
	To overcome the fundamental problem of causal inference—i.e., the inability to observe both potential outcomes for the same individual \citep{Holland1986}—we deploy TabDDPM \citep{Kotelnikov2023}. This Synthetic Data Generation (SDG) architecture learns the complex, high-dimensional joint distribution of the empirical data to synthesize high-fidelity digital twins. Integrated into a T-Learner architecture \citep{Kunzel2019}, this approach allows us to non-parametrically estimate the complete counterfactual manifold for the entire population, bypassing the limitations of traditional propensity score matching \citep{KingNielsen2019} and enabling distributional analysis (specifically, Quantile Treatment Effects) of the most vulnerable EoL cases.
	
	\section{Methodology}
	
	Data were analyzed using \texttt{R} statistical software, while the entire SDG procedure of digital twin generation and evaluation was made via \texttt{Python}. To ensure reproducibility, the full analytical scripts are available at \url{https://github.com/pietrograssi-unifi/Broken-Shield-of-European-PC}.
	
	\subsection{Data Source and Sample Definition}
	
	The empirical foundation of this study relies on SHARE, a cross-national panel database containing harmonized micro-data \citep{BorschSupan2013}. To isolate the EoL phase, we extracted the retrospective EoL (\texttt{xt}) modules from Waves 7 to 9 (2017--2021) across Europe (Israel was excluded due to its distinct institutional context), which record information, such as care utilization and financial outlays, during the patient's final twelve months, as reported by designated proxy respondents. 
	
	The treatment assignment distinguishes between patients who accessed palliative care in the last four weeks ($A=1$) and a specific control group of patients receiving standard care ($A=0$) but whose reason of not undergoing PC treatment depended on availability ($\texttt{xt754\_}=\mbox{2}$) or expensiveness ($\texttt{xt754\_}=\mbox{3}$). It is important to note that this specific SHARE indicator aggregates both home-based palliative care and institutional hospice care into a single binary variable, precluding the structural separation of these two settings. Furthermore, while modern clinical guidelines advocate for the early integration of palliative care well before the terminal phase \citep{Radbruch2020}, our treatment indicator is strictly bound to this final four-week temporal window. To ensure baseline comparability, the \texttt{xt} records were deterministically linked to the patients' socio-demographic and economic profiles from their last regular interview prior to death. 
	
	To maximize statistical power and avoid the severe selection bias inherent in listwise deletion, missing values were addressed via Multivariate Imputation by Chained Equations (MICE) \citep{VanBuuren2011}. Prior to imputation, survey-specific missingness codes were standardized to true missing observations. A pre-imputation audit revealed excellent data completeness for core clinical covariates and direct OOP costs (1.5\%, from \texttt{xt119}). Overall missingness peaked at 35.5\% for the number of children and 28.4\% for the quantification of informal caregiving hours. Rather than relying on arbitrary cutoff thresholds, the application of MICE is justified by methodological evidence demonstrating that multiple imputation yields unbiased estimates even at substantially higher proportions of missingness, provided the Missing At Random (MAR) assumption is plausible \citep{MadleyDowd2019}. Specifically, in strict alignment with current methodological guidelines, we evaluated imputation performance using the Fraction of Missing Information (FMI) rather than the raw proportion of missing data \citep{MadleyDowd2019}. Post-imputation diagnostics across our core analytical models revealed a low average FMI of 15.00\%. This confirms that the rich set of auxiliary variables included in our chained equations successfully captured the underlying missingness mechanism, effectively minimizing between-imputation variance and preserving high statistical efficiency.
	
	To preserve the zero-inflated distribution of caregiving hours and prevent the extrapolation of impossible values, the imputation was performed strictly using Predictive Mean Matching. By conditioning the imputation on a rich vector of fully observed covariates—including the patient's ADL dependency score, wealth quartile, and clinical trajectory—we satisfied the Missing At Random assumption, ensuring a complete and unbiased empirical matrix for the subsequent TabDDPM generative training.
	
	\subsection{Monetization of the Net Total Economic Burden}
	
	To explicitly test the comprehensive relief versus cost-shifting hypotheses, the economic burden was mapped along two primary dimensions: direct liquid toxicity and the shadow cost of time poverty. The former is operationalized as the total out-of-pocket (\texttt{xt119}) healthcare expenditures incurred by the patient during the last twelve months of life (extracted and harmonized into Euros).
	
	Given the categorical nature of the SHARE survey regarding informal care, the frequency of help received (from \texttt{xt024} and \texttt{xt025}) was converted into continuous annual hours through a stochastic smoothing procedure, bounding the physiological maximum of informal caregiving at 16 active hours per day (5,840 hours per year). This specific ceiling is an established methodological standard in health economics, designed to prevent the biological impossibility of 24-hour continuous active care by accounting for the absolute minimum requirement of eight hours for caregiver sleep and basic self-maintenance \citep{Hayman2001, Boogaard2019}. We monetized this informal care burden using the Proxy Good Method \citep{VanDenBerg2004}. Specifically, the estimated caregiving hours were multiplied by the average hourly labor cost of the broader economic sector (NACE Rev. 2, B-S\_X\_O) for the year 2020 in the patient's country of residence. To guarantee exact cross-country comparability and adjust for local living costs, all monetary dimensions were subsequently standardized using Eurostat's dynamic Purchasing Power Parities (PPP) for actual individual consumption \citep{Eurostat2021}. We decided to employ the standard PPP and not the healthcare-specific one beacuse our goal is to measure the impact on households' daily-life purchasing power.
	
	The total economic burden ($Y_i$), expressed in Purchasing Power Standard (PPS) Euros for individual $i$ in country $c$, is thus formalized as:
	$$ Y_i = \underbrace{\frac{OOP_i}{PPP_c}}_{\text{Liquid Toxicity}} + \underbrace{\frac{Hours_i \times Wage_c}{PPP_c}}_{\text{Time Poverty}} $$
	
	\subsection{Causal Identification Framework}
	
	To move beyond observational correlations, we ground our estimation in the Potential Outcomes framework. Let $A_i \in \{0, 1\}$ denote the binary treatment of receiving PC in the last 4 weeks of life, and $Y_i(a)$ the potential economic burden under treatment status $a$. Our primary estimands of interest are the Conditional Average Treatment Effect (CATE), $\mathbb{E}[Y(1) - Y(0) \mid X]$, and the Multivariate Quantile Treatment Effect (QTE), $Q_{\tau}(Y(1) - Y(0) \mid X)$. These targets allow us to unmask both the structural drivers of inequality and the tail-risk failures of the welfare state. Identifying these causal parameters from observational data strictly requires satisfying core assumptions, which we operationalize through our Directed Acyclic Graph (DAG) in Figure \ref{fig:causal_dag}.
	
	\begin{figure}[htbp]
		\centering
		\resizebox{\columnwidth}{!}{
		\begin{tikzpicture}[>=stealth, thick]
			\node[draw, circle, fill=blue!10, minimum size=1cm] (A) at (0,0) {$A$};
			\node[draw, circle, fill=blue!10, minimum size=1cm] (Y) at (7,0) {$Y$};
			
			\node[draw, rectangle, rounded corners, fill=gray!10, align=center] (X) at (3.5, 2.5) {
				\textbf{Observed Covariates $X$} \\ 
				\scriptsize \textbf{Clinical:} Cause of Death, EoL ADL Score, Comorbidities \\ 
				\scriptsize \textbf{Socio-Econ-Demo:} Age, Gender, Marital Status, Wealth, \\
				\scriptsize Homeownership, Subjective Financial Distress \\
				\scriptsize \textbf{Institutional:} Welfare Regime \quad \textbf{Macro-Shock:} COVID-19
			};
			
			\node[draw, rectangle, dashed, rounded corners, fill=red!10, align=center] (U) at (3.5, -2.5) {
				\textbf{Unobserved Confounders $U$} \\ 
				\footnotesize (e.g., hidden family networks, psychological resilience)
			};
			
			\draw[->, ultra thick, draw=blue!60!black] (A) -- (Y);
			\draw[->] (X) -- (A);
			\draw[->] (X) -- (Y);
			\draw[->, dashed, draw=red] (U) -- (A);
			\draw[->, dashed, draw=red] (U) -- (Y);
			
			\draw[<->, dashed, draw=gray, bend right=45] (X.west) to node[midway, left, text=gray] {} (U.west);
		\end{tikzpicture}}
		\caption{Causal Directed Acyclic Graph (DAG) illustrating the identification strategy.}
		\label{fig:causal_dag}
	\end{figure}
	
	\begin{enumerate}
		\item Consistency ($Y = Y(A)$): The SHARE indicator aggregates both home and hospice care. We satisfy this assumption by explicitly defining our estimand not as a narrow clinical protocol, but as the composite state-intervention pathway at the EoL.
		
		\item Positivity ($0 < P(A=1 \mid X) < 1$): The extreme non-linearities and fat-tailed distributions of healthcare costs mechanically destroy the common support required by classical matching estimators. We resolve this algorithmic bottleneck via Synthetic Data Generation, detailed in the subsequent section, which restores the common support.
		
		\item Conditional Exchangeability ($Y(a) \perp A \mid X$): By adjusting for the massive socio-demographic and clinical vector $X$, we block the primary back-door paths (Figure \ref{fig:causal_dag}). However, unobserved confounding ($U$) invariably remains in survey data. We mathematically bound the maximum destructive power of $U$ using the Cinelli-Hazlett's framework \citep{Cinelli2020} (see \ref{app:Cinelli}).
		
		\item No Interference (SUTVA): In capacity-constrained European health systems, one patient accessing a hospice bed might theoretically displace another, violating the assumption of independent potential outcomes. We acknowledge these macro-level supply spillovers and proactively account for the resulting intra-cluster correlation by adjusting our inference via CR2 country-level cluster-robust standard errors.
	\end{enumerate}
	
	\subsection{Causal AI and the Digital Twin Manifold}
	
	To operationalize the potential outcomes framework and satisfy the Positivity assumption, we implemented TabDDPM \citep{Kotelnikov2023}. Rather than forcing baseline balance through classical distance metrics, this generative architecture learns the exact, high-dimensional joint probability distribution of the empirical data $P(X, Y, A)$. This approach allows us to natively accommodate the severe non-linearities and extreme skewness of the economic outcomes, directly synthesizing the missing control equivalent for each patient.
	
	Prior to diffusion, extreme monetary covariates were mapped into a continuous latent space using inverse hyperbolic sine ($\text{arcsinh}$) for household net worth and logarithmic transformations ($\log(1+x)$) for OOP costs, stabilizing the Markovian noise addition process. The model was trained for \num{3000} epochs to synthesize a high-fidelity counterfactual manifold of \num{30000} digital twins. The structural integrity of this synthetic cohort was validated through the FEST framework \citep{niu2025fest}, which comprises metrics on fidelity and overfitting.
	
	\subsection{T-Learner Architecture and ATE Estimation}
	
	Using the synthetic manifold, we implemented a non-parametric Two-Learner (T-Learner) setup \citep{Kunzel2019}. We trained two independent random forest regressors (optimized with 100 estimators and a maximum depth of 15) to map the response surfaces under both the treated ($A=1$) and untreated ($A=0$) potential outcomes:
	$$ \widehat{\mu}_1(X) = \mathbb{E}[Y(1) \mid X,\ A=1] $$
	$$ \widehat{\mu}_0(X) = \mathbb{E}[Y(0) \mid X,\ A=0] $$
	
	Applying G-computation, we passed the original empirical cohort through both models to predict the expected economic burden under each scenario for every real patient. This procedure effectively neutralizes the severe baseline imbalances observed in the raw empirical data—where untreated patients were disproportionately concentrated in lower wealth quartiles and underfunded Eastern regimes (Table \ref{tab:empirical_cohort})—by generating a universally balanced digital twin cohort (Table \ref{tab:ate_cohort}). The Individual Treatment Effect (ITE) was derived as $\widehat{\Delta}_i = \widehat{Y}_{i}(1) - \widehat{Y}_{i}(0)$. To mitigate the artificial variance introduced by algorithmic boundary predictions on highly skewed data, the resulting ITE distributions were winsorized at the 1st and 99th percentiles. The global ATE was then computed applying a Bias-Corrected and Accelerated (BCa) bootstrap with \num{1000} iterations to correctly account for non-Gaussian asymmetries \citep{Efron1987}.
	
	\subsection{Econometric Modeling of Institutional and Clinical Inequality}
	
	To operationalize the concepts of disease-based and institutional inequality, we investigated the structural drivers of the causal shift. The Conditional Average Treatment Effect (CATE) was estimated via Ordinary Least Squares (OLS) regression on the individual causal deltas. The model was independently fitted for the three primary outcomes (OOP expenditures, caregiving hours, and net total burden) using the following specification:
	$$ \widehat{\Delta}_i = \beta_0 + \beta_1\text{WR}_i + \beta_2\text{CoD}_i + X_i^\top \gamma + \epsilon_i$$
	
	In this equation, $\beta_1$ and $\beta_2$ are the coefficients of primary interest, isolating the institutional and clinical penalties on the treatment effect. The term $\text{WR}_i$ refers to the welfare regime, while $\text{CoD}_i$ is the cause of death. Furthermore, the term $X_i$ represents a vector of socio-demographic and clinical covariates. Clinically, we control for the exact EoL functional dependency (ADL score, \texttt{xt020}) and the baseline number of chronic morbidities (\texttt{ph006}). Economically, we control for the household's net wealth quartile (\texttt{hnetw}), formal homeownership status (\texttt{otrf} from \texttt{ho002}), and pre-existing subjective financial distress (\texttt{fdistress} from \texttt{co007}), measured as the household's self-reported ability to make ends meet. By extracting this socio-economic profile from the last regular survey wave prior to the patient's death, we proactively neutralize the risk of reverse causality, ensuring that our models capture the effect of baseline structural vulnerability rather than the ex-post distress caused by the EoL OOP shock itself. Finally, a binary fixed effect for the COVID-19 pandemic period (Wave 9) is included to mathematically absorb the extreme macroeconomic friction and price volatility of the 2020--2021 lockdowns, isolating deeply rooted systemic inequalities from the exogenous pandemic shock.
	
	Since the survey respondents are naturally nested within 27 distinct European health systems, assuming independent and identically distributed error terms would ignore intra-country correlation, leading to severely underestimated variance and inflated Type I error rates. However, standard cluster-robust standard errors (like the Liang-Zeger CR0 estimator) are asymptotically biased downward when the number of clusters is small (typically less than 40) or when cluster sizes are unbalanced \citep{CameronMiller2015}. 
	
	To resolve this finite-sample bias, we computed the variance-covariance matrix using the CR2 cluster-robust adjustment \citep{BellMcCaffrey2002}. Mathematically, rather than using the raw OLS residuals ($\widehat{\epsilon}_i$), the CR2 estimator inflates them by scaling each residual using the diagonal elements $h_{i,i}$ of the projection matrix (the hat matrix). By applying the adjustment factor $(1 - h_{i,i})^{-1/2}$, the CR2 method explicitly corrects for the mechanical tendency of OLS to fit data points with high leverage too closely, thereby guaranteeing reliable and unbiased hypothesis testing even with only 27 country-level clusters.
	
	Furthermore, to test the ``broken shield'' hypothesis, we estimated Multivariate Quantile Treatment Effects (QTE) to evaluate whether the protective capacity of the intervention collapses for the most critical clinical cases \citep{Koenker1978}. QTE avoids minimizing squared residuals and instead estimates the effect across specific severity percentiles ($\tau \in \{0.50, 0.75, 0.90\}$) using an asymmetric absolute loss function:
	$$ \min_{\beta_\tau} \sum_{i=1}^{N} \rho_\tau (\widehat{\Delta}_i - X_i^\top \beta_\tau)$$ 
	where $\rho_\tau(u) = u(\tau - \mathds{1}(u < 0)) $. To bypass singular design matrix errors and non-positive densities caused by exact ties in categorical dummy variables—a known mechanical limitation of quantile estimators on survey data \citep{Machado2005}—we applied uniform micro-jittering ($[-10^{-4}, 10^{-4}]$) to the continuous outcome prior to applying $xy$-pair bootstrap resampling for standard error computation.
	
	Finally, to address the vulnerability of observational causal inference to hidden biases, we implemented the state-of-the-art robustness framework developed by \citet{Cinelli2020} (see Appendix \ref{app:Cinelli}). By computing partial $R^2$ contour bounds and robustness values, this framework mathematically quantifies the explanatory power that an unobserved, hypothetical confounder (e.g., unmeasured pre-existing psychological resilience or hidden family network dynamics) would need to possess to successfully invalidate our significant structural estimates—specifically the massive penalty associated with non-cancer trajectories—benchmarking it against predictive mechanical covariates such as the EoL functional dependency (ADL) score.
	
	\section{Results}
	
	\subsection{Generative Fidelity and Privacy Assessment}
	
\begin{figure}[htbp]
		\centering
		\includegraphics[width=\columnwidth]{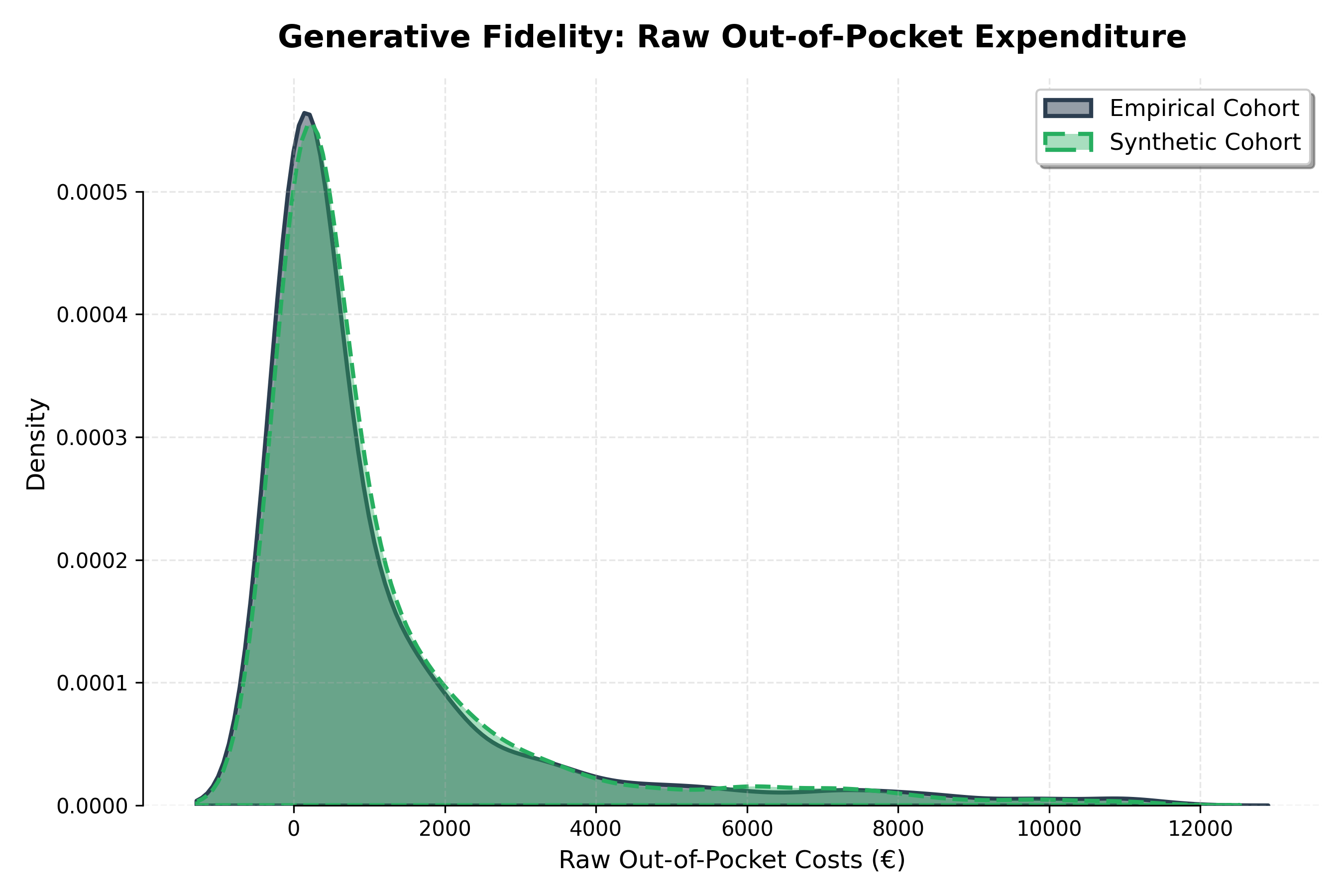}
		\caption{Generative Fidelity of TabDDPM: Empirical vs. Synthetic distribution of raw OOP expenditures, demonstrating accurate replication of zero-inflation and extreme right-skewness.}
		\label{fig:fest_fidelity}
	\end{figure}

	Before analyzing the causal effects, the quality of the synthetic counterfactual manifold generated via TabDDPM was evaluated using the FEST framework. The generative model achieved a Kolmogorov-Smirnov (KS) Average Score of 0.869 and a Mean Absolute Percentage Error (MAPE) of 0.161, demonstrating exceptional distributional overlap with the empirical SHARE data. Crucially, the model successfully replicated the complex multivariate relationships of the original data, evidenced by a Frobenius Correlation Difference of 0.954, while perfectly maintaining the zero-inflation and fat-tailed asymptotes characteristic of OOP healthcare expenditures (Figure \ref{fig:fest_fidelity}). 
	
	Regarding disclosure risk and patient privacy, the Distance to Closest Record reached 0.284, confirming that the synthesized digital twins are sufficiently distant from their real counterparts and are not mere clones of the original observations \citep{Kotelnikov2023}. Furthermore, the Adversarial Accuracy score was 0.697. These metrics certify that the generative architecture did not memorize the training data, operating well within safe privacy thresholds for sensitive medical records \citep{Yale2020}. Demographic and baseline characteristics for both the empirical and the universally balanced ATE cohorts are detailed in Tables 1a and 1b (available in the Appendix).
	
	\subsection{Distributional Causal Shifts}
	
\begin{figure}[htbp]
		\centering
		\begin{minipage}{\columnwidth}
			\centering
			\includegraphics[width=\linewidth]{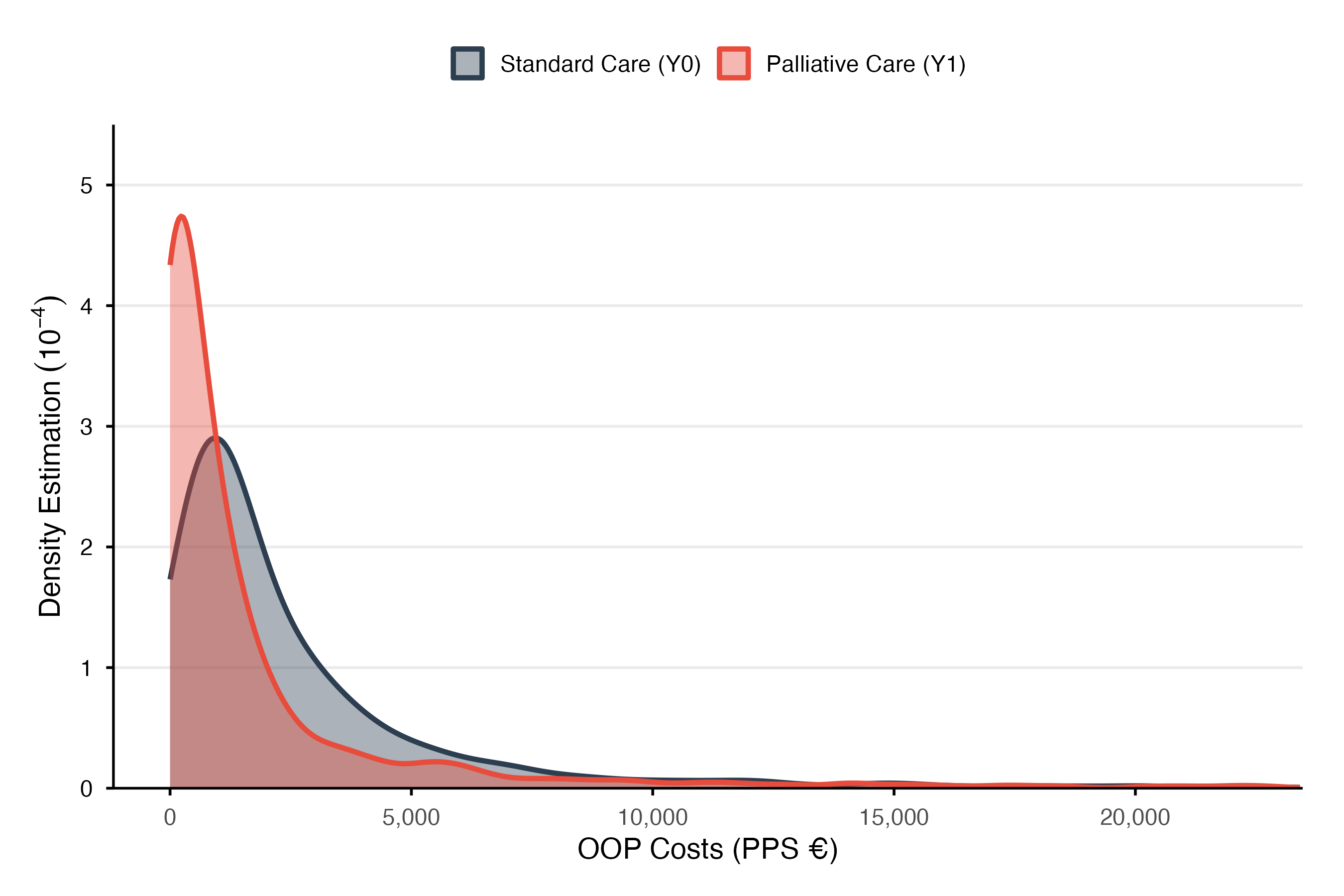}
			\caption{Causal Shift in Liquid Toxicity (OOP Costs in PPS €).}
			\label{fig:shift_oop}
		\end{minipage}\hfill
		\begin{minipage}{\columnwidth}
			\centering
			\includegraphics[width=\linewidth]{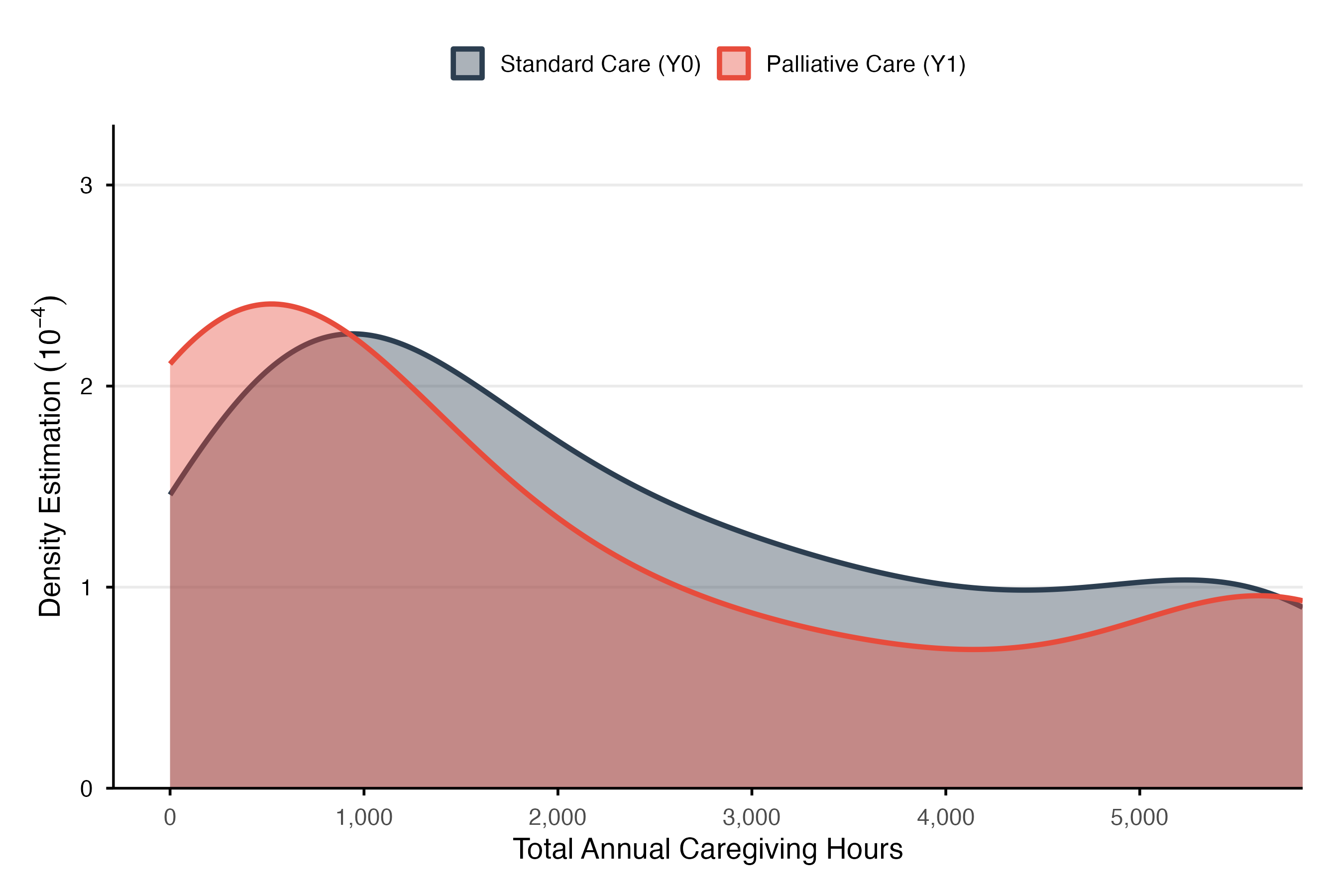}
			\caption{Causal Shift in Labor Substitution (Informal Caregiving Hours).}
			\label{fig:shift_hours}
		\end{minipage}
	\end{figure}

\begin{figure}[htbp]
		\centering
		\includegraphics[width=\columnwidth]{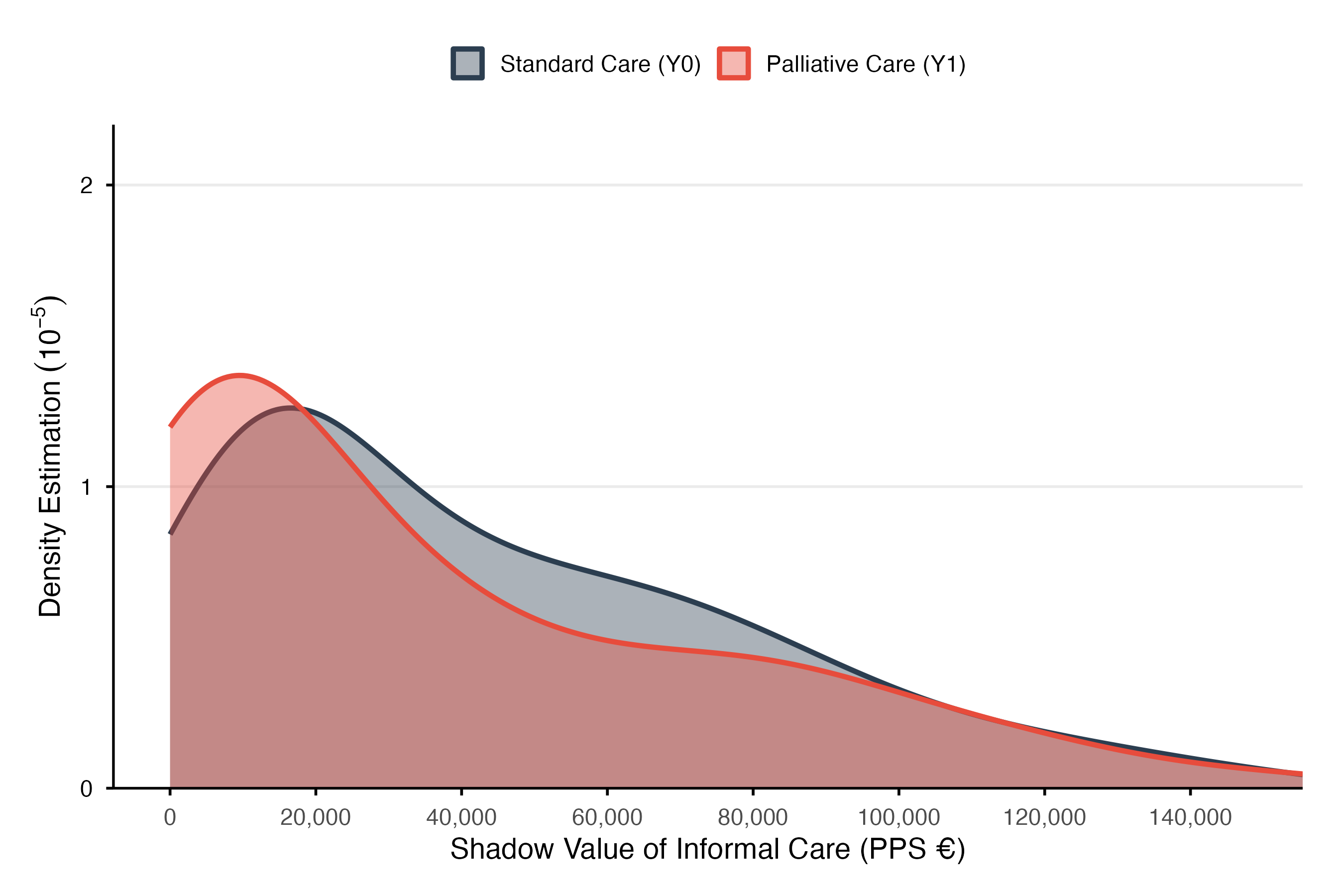}
		\caption{Causal Shift in the Monetized Value of Informal Care (PPS €).}
		\label{fig:shift_informal}
	\end{figure}

	The application of the non-parametric T-Learner on the full European cohort reveals a macroscopic shift in the economic burden. Figures \ref{fig:shift_oop}, \ref{fig:shift_hours}, and \ref{fig:shift_informal} illustrate the Kernel Density Estimates for the expected potential outcomes under standard care ($Y_0$) versus PC ($Y_1$).
	
	Contrary to the strict cost-shifting hypothesis, the introduction of PC heavily concentrates the mass of the distribution near zero across all dimensions of the economic burden, effectively truncating the main body of the right tail. Figure \ref{fig:shift_oop} shows a marked concentration of OOP costs near zero under the $Y_1$ scenario, indicating substantial relief from direct liquid toxicity for the vast majority of the population. More notably, this financial protection does not systematically occur at the expense of family caregivers: Figure \ref{fig:shift_hours} and Figure \ref{fig:shift_informal} demonstrate that PC also significantly reduces the aggregate volume of informal caregiving hours and its corresponding monetized shadow value across the bulk of the distribution.
	
	However, a closer inspection of the extreme right margins reveals a crucial nuance. While PC successfully suppresses the burden for most of the population, at the absolute tail, the PC density curve ($Y_1$) slightly overtakes standard care ($Y_0$). This visual anomaly indicates that for a small fraction of extreme cases, the transition to PC may paradoxically extend or exacerbate extreme time poverty and OOP expenditures. This subtle distributional persistence foreshadows the structural inequalities unmasked by the QTE analysis in the following sections.
	
	\subsection{Conditional Treatment Effects}
	
	As detailed in Table \ref{tab:cate_models}, the baseline global effect for the absolute reference group (intercept) shows a muted and statistically non-significant effect regarding both labor substitution ($+88.6$ caregiving hours) and the net economic burden ($+2,329.1$ PPS €). This indicates that within our highly granular covariate architecture, the theoretical baseline protection is largely absorbed and effectively distributed across the specific underlying clinical and socioeconomic vulnerabilities.
	
	\begin{table}[htbp]
		\centering
		\caption{Multivariate Analysis of CATE (CR2 Cluster-Robust Standard Errors)}
		\label{tab:cate_models}
		\resizebox{\columnwidth}{!}{%
			\begin{tabular}{lccc}
				\toprule
				\textbf{Variable} & \textbf{$\Delta$ OOP Costs} & \textbf{$\Delta$ Care} & \textbf{$\Delta$ Net Burden} \\
				& \textbf{(PPS €)} & \textbf{(Hours)} & \textbf{(PPS €)} \\ 
				\midrule
				Intercept & $-222.513$ & $88.614$ & $2329.147$ \\
				CoD: Cancer & Ref. & Ref. & Ref.\\
				CoD: Organ Failure & $322.751$ & $69.106$ & $2696.757$ \\
				CoD: Other & $506.084^{+}$ & $420.784^{**}$ & $8698.318^{**}$ \\
				Welfare: Continental & Ref. & Ref. & Ref.\\
				Welfare: Eastern & $-355.130$ & $267.707$ & $8172.445$ \\
				Welfare: Nordic & $-393.753$ & $560.101$ & $12690.697$ \\
				Welfare: Southern & $-652.155$ & $260.249$ & $7596.515$ \\
				Wealth: Q1 (Poorest) & Ref. & Ref. & Ref.\\
				Wealth: Q2 & $-709.514$ & $-57.359$ & $-2924.443$ \\
				Wealth: Q3 & $-1055.310^{*}$ & $-28.691$ & $-1245.440$ \\
				Wealth: Q4 (Richest) & $-698.411$ & $218.850$ & $3679.213$ \\
				Age & $7.514$ & $18.232^{**}$ & $258.468^{***}$ \\
				Gender: Male & $-347.017$ & $137.986^{+}$ & $1599.991$ \\
				Marital Status: Single & $-498.053^{*}$ & $168.174$ & $1289.431$ \\
				EoL ADL Score & $-135.245^{*}$ & $-52.486$ & $-1189.509$ \\
				Home Owner: Yes & $-474.908$ & $206.565$ & $3302.215$ \\
				Financial Distress & $-128.529$ & $219.782^{**}$ & $4771.748^{**}$ \\
				Comorbidities & $164.717^{*}$ & $33.682$ & $946.402$ \\
				Pandemic: During COVID-19 & $1595.041^{*}$ & $-25.751$ & $1509.303$ \\
				\midrule
				Number of Observations & 2192 & 2192 & 2192 \\
				$R^2$ & $0.057$ & $0.061$ & $0.075$ \\
				\bottomrule
				\multicolumn{4}{l}{\footnotesize \textit{Note: $^{+} p<0.1$, $^{*} p<0.05$, $^{**} p<0.01$, $^{***} p<0.001$. Standard errors clustered by country.}}\\
			\end{tabular}%
		}
	\end{table}
	
	Institutionally, while residing in an Eastern European or a Southern welfare regime shows large positive point estimates for the net total burden penalty ($+8,172.4$ PPS € and $+7,596.5$ PPS €, respectively), these regional effects completely lack formal statistical significance at the conditional average level. This reveals that average institutional inequality is likely to be mediated by underlying household financial vulnerability. Conversely, clinically, the penalties for non-cancer terminal trajectories explicitly emerge as massive structural drivers: the ``other'' cause of death (e.g., severe frailty and dementia) imposes a highly significant penalty of $+8,698.3$ PPS € ($p<0.01$) on the net total burden, actively driven by simultaneous surges in both direct OOP costs ($+506.1$ PPS €, $p<0.1$) and labor substitution ($+420.8$ informal caregiving hours, $p<0.01$). As detailed in Appendix \ref{app:Cinelli}, this clinical penalty mathematically resists the potential presence of unobserved confounders even when benchmarked against severe physical dependency. This crucial finding suggests that clinical vulnerabilities dictate profound inequality even at the average level, foreshadowing extreme tail events that will be exposed by the QTE models.
	
	Socio-demographic profiles critically modulate this shock, validating our theoretical expectations regarding intra-household rigidities. Lacking a spousal safety net (single) significantly enhances the reduction of direct OOP spending ($-498.1$ PPS €, $p<0.05$), and a similar localized reduction is observed for patients in the third wealth quartile ($-1,055.3$ PPS €, $p<0.05$), although neither translates into a significant comprehensive relief on the net burden. Reflecting the theorized rigidities of gendered labor market participation, male patients generate a systematically higher marginal demand for informal care ($+138.0$ hours, $p<0.1$) compared to females. Furthermore, demographic vulnerability is strongly compounded by age, which significantly increases both the care deficit ($+18.2$ hours per year, $p<0.01$) and the net burden ($+258.5$ PPS €, $p<0.001$). As expected, the covariate for subjective financial distress acts as a massive driver across the models, increasing labor substitution by $+219.8$ hours ($p<0.01$) and the net burden by $+4,771.7$ PPS € ($p<0.01$) per unit of distress. Similarly, baseline clinical comorbidities (chronic) directly inflate OOP costs ($+164.7$ PPS €, $p<0.05$). Interestingly, a higher functional dependency at the end of life (EoL ADL Score) shows a marginal substitution dynamic, slightly reducing direct OOP spending ($-135.2$ PPS €, $p<0.05$).
	
	Finally, our modeling architecture explicitly isolates these structural findings from the exogenous macroeconomic shock of the 2020--2021 lockdowns through the inclusion of a COVID-19 fixed effect. While the pandemic era exerted a statistically significant inflationary pressure on direct OOP spending ($+1,595.0$ PPS €, $p<0.05$), the persistence of the specific clinical and demographic penalties confirms they are deeply rooted structural failures rather than transient artifacts of pandemic-induced healthcare stress. Detailed temporal sensitivity analyses are further provided in Appendix \ref{app:COVID}.
	
	\begin{figure}[htbp]
		\centering
		\includegraphics[width=\columnwidth]{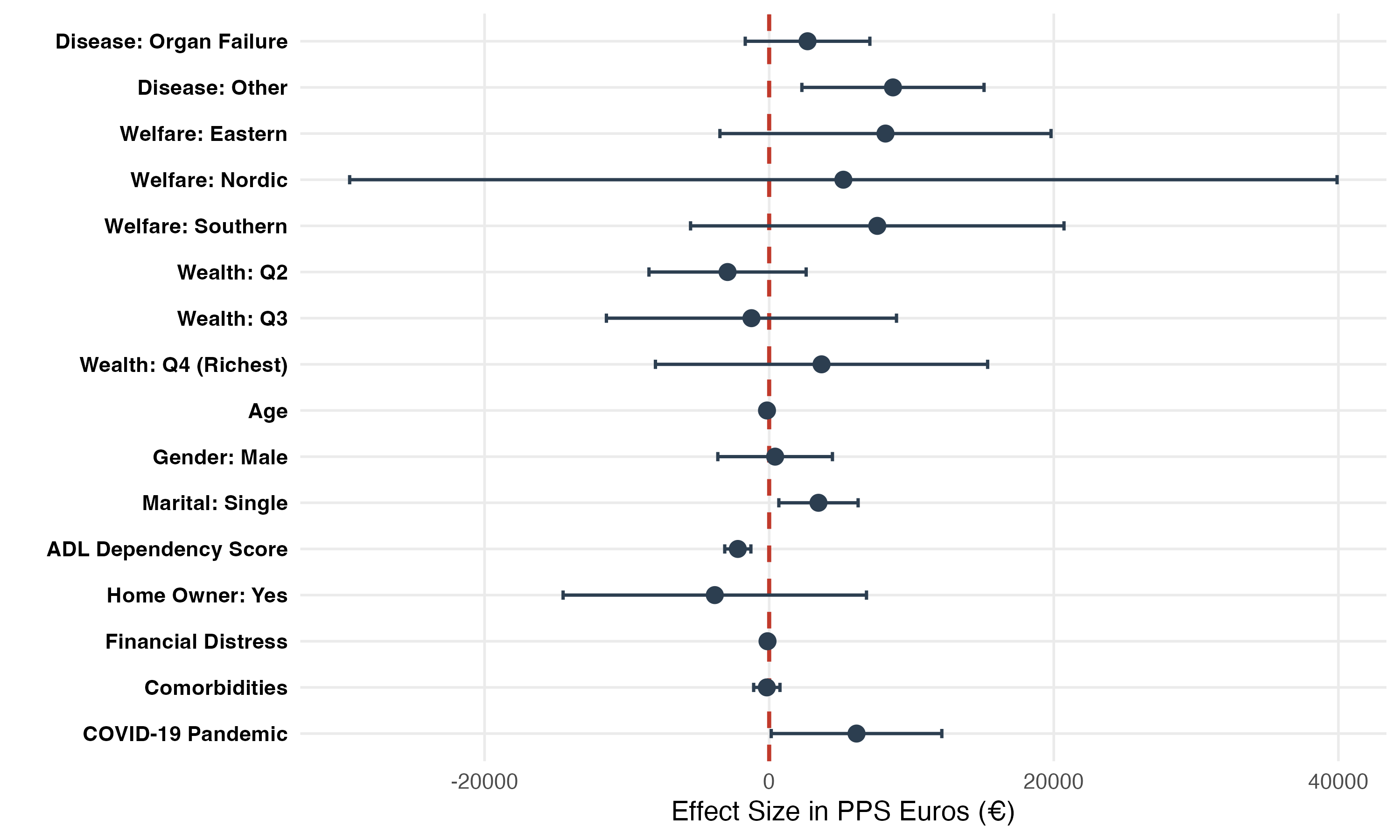}
		\caption{Marginal CATE Forest Plot displaying the socio-institutional heterogeneity of welfare loss.}
		\label{fig:forest_plot}
	\end{figure}
	
	\begin{figure}[htbp]
		\centering
		\includegraphics[width=\columnwidth]{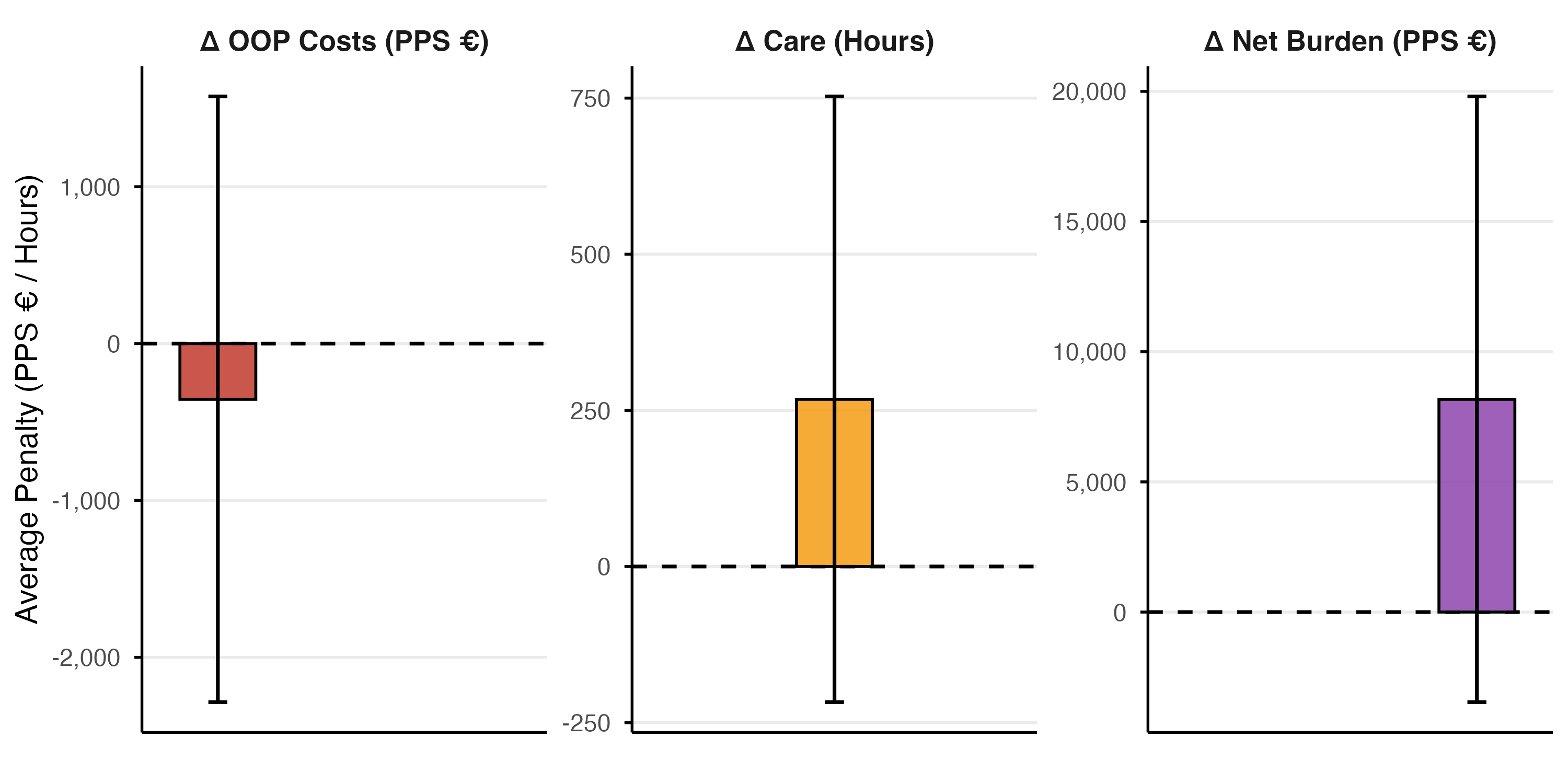}
		\caption{Institutional inequality: Average Treatment Effect penalties for transitioning to palliative care in Eastern European regimes across multiple economic dimensions.}
		\label{fig:cate_eastern}
	\end{figure}
	
	\subsection{Quantile Treatment Effects}
	
	Turning to the distributional extremes (Table \ref{tab:qte_models}), at the median (50th percentile), the baseline net burden intercept stands at $-4,327.0$ PPS € (non-significant). However, analyzing the tails reveals a stark deterioration of the palliative shield: the trajectory escalates into a penalty of $+7,174.3$ PPS € at the 75th percentile and explodes into a massive $+39,303.6$ PPS € ($p<0.05$) penalty at the 90th percentile. This trajectory confirms that for the 10\% of households facing the most grueling realities, the theoretical protection of PC is severely strained.
	
\begin{figure}[htbp]
		\centering
		\includegraphics[width=\columnwidth]{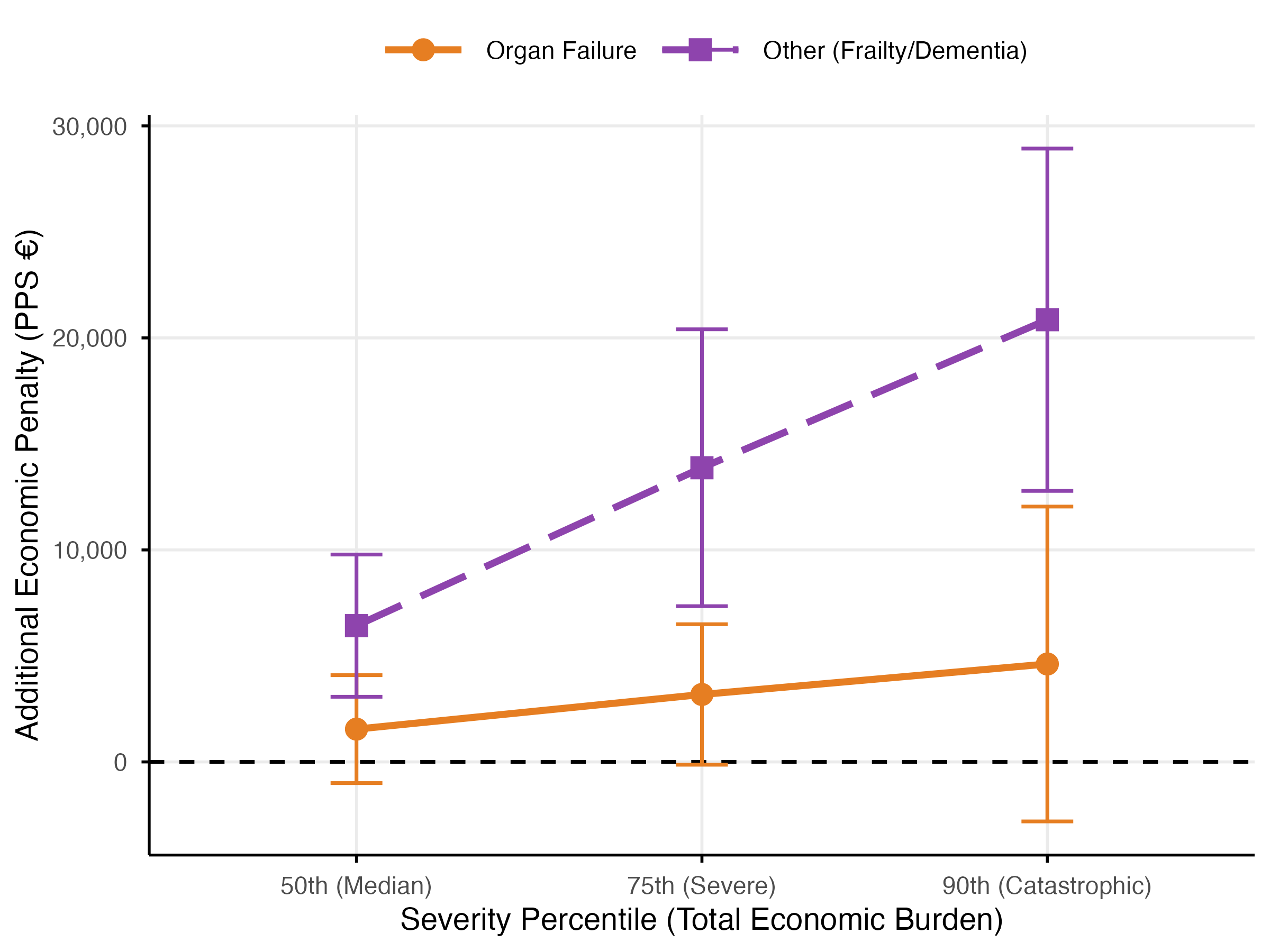}
		\caption{QTE Dynamics of Clinical inequality: The economic penalty for non-cancer trajectories escalates dramatically at the right tail of the burden distribution.}
		\label{fig:qte_clinical}
	\end{figure}

	Most strikingly, clinical penalties for non-cancer trajectories escalate in the upper quantiles (Figure \ref{fig:qte_clinical}). The ``other'' cause of death drives an economic harm scaling from $+6,427.3$ PPS € at the median ($p<0.001$) to an extreme $+20,857.4$ PPS € at the 90th percentile ($p<0.001$). Organ failure similarly imposes a $+3,180.6$ PPS € penalty at the severe phase ($p<0.1$), confirming the profound economic vulnerabilities associated with prolonged and unpredictable dying processes. As expected, the functional dependency index (EoL ADL Score) acts as a mechanical control variable, absorbing the endogenous costs dictated by physical loss. This compounding effect scales from $+449.1$ PPS € ($p<0.1$) at the median to $+3,037.2$ PPS € ($p<0.001$) at the tail for every additional point of functional loss, effectively isolating the remaining institutional and socio-demographic penalties.
	
	\begin{table}[htbp]
		\centering
		\caption{Multivariate Quantile Treatment Effects on Net Economic Burden}
		\label{tab:qte_models}
		\resizebox{\columnwidth}{!}{%
			\begin{tabular}{lccc}
				\toprule
				\textbf{Variable} & \textbf{Median} & \textbf{Severe} & \textbf{Extreme} \\
				& \textbf{(50th)} & \textbf{(75th)} & \textbf{(90th)} \\ 
				\midrule
				Intercept & $-4327.042$ & $7174.307$ & $39303.613^{*}$ \\
				CoD: Cancer & Ref. & Ref. & Ref.\\
				CoD: Organ Failure & $1546.146$ & $3180.589^{+}$ & $4618.472$ \\
				CoD: Other & $6427.285^{***}$ & $13875.262^{***}$ & $20857.436^{***}$ \\
				Welfare: Continental & Ref. & Ref. & Ref.\\
				Welfare: Eastern & $6535.957^{*}$ & $4142.604^{+}$ & $-6122.903$ \\
				Welfare: Nordic & $3907.387^{+}$ & $5084.651$ & $908.917$ \\
				Welfare: Southern & $5122.318^{*}$ & $1510.476$ & $-5114.727$ \\
				Wealth: Q1 (Poorest)& Ref. & Ref. & Ref.\\
				Wealth: Q2 & $-617.618$ & $-4562.943^{*}$ & $-7446.526$ \\
				Wealth: Q3 & $-1672.647$ & $-3408.307$ & $-6866.276$ \\
				Wealth: Q4 (Richest) & $1786.032$ & $1134.112$ & $1249.856$ \\
				Age & $5.109$ & $24.331$ & $-29.284$ \\
				Gender: Male & $2292.016^{*}$ & $3600.676^{+}$ & $8068.740^{*}$ \\
				Marital Status: Single & $1686.038$ & $166.726$ & $-5679.524$ \\
				EoL ADL Score & $449.125^{+}$ & $2289.479^{***}$ & $3037.195^{***}$ \\
				Home Owner: Yes & $382.464$ & $-2026.046$ & $-11499.780$ \\
				Financial Distress & $1292.365^{*}$ & $2465.045^{*}$ & $6100.835^{+}$ \\
				Comorbidities & $-155.617$ & $383.181$ & $1391.134$ \\
				Pandemic: During COVID-19 & $-36.634$ & $4376.108^{*}$ & $10738.835^{+}$ \\
				\midrule
				Number of Observations & 2192 & 2192 & 2192 \\
				$R^2$ & $0.015$ & $-0.379$ & $-1.954$ \\
				\bottomrule
				\multicolumn{4}{l}{\footnotesize \textit{Note: $^{+} p<0.1$, $^{*} p<0.05$, $^{**} p<0.01$, $^{***} p<0.001$. Bootstrapped $xy$-pair standard errors.}} \\
			\end{tabular}%
		}
	\end{table}
	
	Validating our theoretical expectations, individual socio-demographics dictate the distributional shock. While lacking a spousal safety net (single) suggests a directional penalty at the median ($+1,686.0$ PPS €), it lacks formal statistical significance. Conversely, male patients impose significantly higher costs across the entire distribution, peaking at $+8,068.7$ PPS € at the 90th percentile ($p<0.05$). The compounding shock of subjective financial distress is distinctly tail-driven, inflating the burden by $+1,292.4$ PPS € ($p<0.05$) at the median and $+6,100.8$ PPS € ($p<0.1$) at the extreme tail. While point estimates for the upper wealth brackets consistently show negative coefficients, these effects largely lack formal statistical significance, with the exception of a localized buffering effect for the second quartile ($-4,562.9$ PPS €, $p<0.05$). This dynamic underscores the extreme variance of the right tail, suggesting that high OOP costs are wildly unpredictable and can drain accumulated wealth across almost all income groups.
	
	This massive volatility at the extremes is further illuminated by the COVID-19 fixed effect. While the pandemic shock is statistically non-significant at the median, it explicitly targets the tails, inflating the burden by an additional $+4,376.1$ PPS € at the 75th percentile ($p<0.05$) and $+10,738.8$ PPS € at the 90th percentile ($p<0.1$). This precisely reflects the profound unpredictability and price volatility of shadow care markets during global lockdowns for the most severe cases, underscoring that our core structural findings persist even when controlling for this intense systemic friction.
	
	The institutional penalty presents a complex dynamic across the European landscape, strictly aligning with our comparative welfare hypotheses. Confirming the structural paradox of high-wage, dual-earner economies, the Nordic regime---despite its robust formal domiciliary networks---exhibits a substantial opportunity cost penalty at the median ($+3,907.4$ PPS €, $p<0.1$) relative to the Continental baseline. The Southern regime mirrors this early strain, imposing a $+5,122.3$ PPS € penalty at the median ($p<0.05$). Most notably, the Eastern regime shows a large and highly significant penalty at the median ($+6,535.9$ PPS €, $p<0.05$) and 75th percentile ($+4,142.6$ PPS €, $p<0.1$), but sharply reverses into a negative coefficient at the 90th percentile ($-6,122.9$ PPS €). Although this extreme tail effect lacks formal statistical significance due to massive variance, the trajectory confirms our expectations regarding a potential structural ceiling: households facing economic hardship in underfunded regimes might simply exhaust their spending capacity rather than experiencing genuine economic relief.
	
	\section{Discussion}
	
	\subsection{Rejecting the Cost-Shifting Paradigm and the Fiscal Illusion}
	
	The most prominent macroscopic finding of our causal analysis is the rejection of the strict cost-shifting hypothesis for the general population. Previous critical sociology and health economics literature has raised concerns that the deinstitutionalization of dying—shifting care from hospitals to the community—might inherently transfer the economic burden onto families via unremunerated labor \citep{Gardiner2020, Urwin2021}. 
	
	Our distributional causal shifts prove otherwise, yet with a crucial caveat. When adequately implemented, PC does not merely substitute state-funded hospital beds with unpaid family labor; at a macroeconomic level, it successfully truncates the right tail of out-of-pocket expenditures and caregiving hours. However, as demonstrated by our highly granular CATE models, the baseline ``double shield'' effect is neither automatic nor universal. By controlling for subjective financial distress and specific clinical comorbidities, the baseline protection for the absolute reference group is largely absorbed. This demonstrates that failing to account for unremunerated labor perpetuates a fiscal illusion, where the narrow perspective mandated by regulatory authorities \citep{NICE2025, Hoefman2013} successfully protects robust households but heavily relies on transferring time poverty and liquid toxicity onto those who are already clinically and economically vulnerable \citep{Gardiner2020}. As detailed in Appendix \ref{app:SHADOW}, these macroscopic conclusions regarding the shadow cost of time remain highly robust to extensive sensitivity analyses on the proxy good monetization methodology.
	
	Most importantly, this macroscopic optimism is shattered at the tails of the distribution. As our QTE models reveal, the baseline intercept explodes into a huge $+39,303.6$ PPS € penalty at the 90th percentile. At the extreme 10\% tail, we observe the most prolonged trajectories requiring continuous 24/7 supervision. When dying under standard care, this heavy nursing labor and lodging are generally absorbed by European health coverage \citep{Bekelman2016, French2017}. Conversely, transitioning to the PC pathway exposes the household to structural costs. If care is provided at home, episodic formal visits are insufficient to cover the 24-hour void, forcing families to absorb thousands of hours of intensive informal labor or purchase private shadow care. If care is provided in a hospice, families frequently face substantial OOP co-payments or long-term facility fees that are not fully state-subsidized.
	
	\subsection{Tail-Driven Clinical Inequality and the Failure of Risk-Bearing Markets}
	
	Our models expose a severe architectural flaw in contemporary EoL systems for complex clinical trajectories. Historically, the PC paradigm was engineered around the linear and predictable terminal decline of oncology \citep{Murray2005}. Conversely, patients dying from severe frailty, dementia, and organ failure experience a trajectory of ``prolonged dwindling'' \citep{Etkind2017}. 
	
	Crucially, this disease-based inequality acts as a massive structural driver. Our CATE reveals that the ``other'' cause of death drives a highly significant average penalty of $+8,698.3$ PPS € ($p<0.01$). Furthermore, our QTE models demonstrate that this penalty explodes at the extreme tails of the distribution. The economic harm for non-cancer trajectories escalates violently, with the ``other'' cause of death driving a massive $+20,857.4$ PPS € penalty at the 90th percentile, and organ failure imposing a $+3,180.6$ PPS € penalty at the severe 75th percentile. 
	
	As expected, this structural collapse of the palliative shield is mechanically compounded by the patient's physical deterioration. The EoL ADL Score confirms that functional dependency inevitably scales the economic burden across the distribution. Explaining this dynamic, \citet{Bonsang2008} demonstrates that informal caregiving acts as an effective substitute for formal paid care only at low-to-moderate levels of patient dependency. During extreme clinical deterioration, this substitution effect vanishes; informal and formal care cease to be substitutes and become strict complements. In the context of protracted EoL trajectories, private insurance markets fail to emerge effectively due to the extreme aggregate uncertainty of forecasting survival durations and care costs decades into the future \citep{CostaFont2015}. Deprived of both functional private risk pooling and adequate public LTC entitlements, the household is thrust into the role of the ultimate insurer, absorbing the entirety of this tail risk.
	
	To ensure that this disease-based inequality is not an artifact of omitted variable bias, our robustness analysis utilizing the \citet{Cinelli2020} framework (Appendix \ref{app:Cinelli}) confirms that the non-cancer penalty mathematically resists the potential presence of massively powerful unobserved confounders. Specifically, an unmeasured shock would need to be substantially more powerful than the patient's entire severe physical dependency to overturn the statistical significance of this clinical failure.
	
	\subsection{Labor Market Frictions and Gendered Hysteresis}
	
	The socio-demographic characteristics of the patient critically dictate how households internalize this EoL shock. Lacking a spousal safety net (single) imposes a visible directional penalty on the monetized net burden ($+1,686.0$ PPS € at the median). Severe liquidity constraints effectively depress single patients' direct OOP spending, forcing the terminal caregiving burden entirely onto extended informal networks (e.g., adult children) and triggering a massive surge in time poverty.
	
	Furthermore, our models reveal a profound gendered divergence in how the EoL shock impacts the labor market. Male patients generate a systematically higher demand for informal care with respect to females ($+138.0$ hours in the CATE) and drive severe economic penalties across the distribution, peaking at $+8,068.7$ PPS € at the 90th percentile. Because male caregivers are historically embedded in an ``ideal worker culture''—demanding unwavering, full-time devotion to the workplace—they face severe structural constraints in adjusting their labor supply at the intensive margin \citep{Chung2020, Lott2016}, thus the difference in needs between male and female patients may be a supply problem, rather than an actual demand one, in the sense that—because of rooted gender norms—men may not be as able as women to provide support when their spouses need it.
	
	Conversely, European ``family-friendly'' policies systematically, albeit inadvertently, channel female workers into part-time or flexible employment arrangements \citep{Blau2013}. When the EoL care shock occurs, female caregivers tend to exploit this flexibility, absorbing the massive caregiving demand by further reducing working hours rather than exiting entirely. While this forced flexibility artificially softens the immediate liquid toxicity, it masks a profound macroeconomic scar. Driven by persistent occupational hysteresis, these female caregivers suffer long-term human capital depreciation and reduced pension benefits \citep{Heger2020}. Thus, the lack of institutional domiciliary support triggers immediate labor market ejection for men, and irreversible, hidden financial scarring for women.
	
	\subsection{Institutional Inequality, the Wealth Buffer, and the Pandemic Stress Test}
	
	Institutionally, our results demand a reinterpretation of EoL economics across Europe, revealing distinct mechanisms of welfare loss. In the Nordic regime, the significant penalty observed ($+3,907.4$ PPS € at the median) exposes the friction of a high-wage, dual-earner macroeconomic structure. Because the Nordic baseline relies heavily on state defamilialization, shifting a highly dependent dying patient to the home generates a steep marginal shock. In economies characterized by maximized labor market participation, the forced withdrawal of family members to bridge residual care gaps translates into a massive monetized opportunity cost \citep{Szebehely2018}. Conversely, in the Southern regime, the intense structural reliance on the family drives a $+5,122.3$ PPS € penalty at the median. 
	
	Most notably, subjective financial distress actively absorbs the regional variance in the average CATE estimations, proving that the institutional failure of underfunded Eastern regimes is profoundly mediated by underlying household poverty. This translates into brutal distributional effects: at the median, Eastern families face a massive $+6,535.9$ PPS € penalty, but at the 90th percentile, this sharply reverses into a paradoxical negative QTE coefficient ($-6,122.9$ PPS €). This does not indicate genuine economic relief. Instead, it captures the precise threshold at which households hit an absolute, binding budget constraint—a dynamic conceptually aligned with ``catastrophic health spending'' \citep{Wagstaff2018}. Because families in these regimes cannot substitute formal nursing with their own time, and cannot afford to purchase it privately, they simply run out of resources and cease formal consumption entirely.
	
	We hypothesized that accumulated private wealth would act as a surrogate insurance buffer against these institutional failures. While our QTE models show a localized protective effect for the second wealth quartile ($-4,562.9$ PPS € at the 75th percentile), this buffering capacity largely loses formal statistical significance at the extreme tails. This extreme variance perfectly aligns with the economics of asset depletion \citep{Poterba2017}: large EoL OOP costs are so wildly unpredictable that they can rapidly drain accumulated wealth across almost all income groups.
	
	This vulnerability is further compounded by the structural nature of household assets. Formal homeownership was explicitly included in our covariate matrix because it acts as a definitive marker of established lifestyle, capturing underlying socio-demographic advantages and baseline living standards that transcend mere cash flow \citep{Grundy2001}. However, despite its strong theoretical role in defining a household's structural capacity, this variable failed to yield statistical significance across both the CATE and QTE estimations. While a home represents accumulated wealth, it simultaneously functions as an indispensable physical asset: it provides essential shelter for the family and the fundamental infrastructure required for domiciliary EoL care. Therefore, the inability to leverage this wealth is not merely a matter of market illiquidity, but of strict functional indivisibility. A household cannot liquidate their primary residence to finance immediate OOP medical liabilities without simultaneously destroying the patient's care setting or forcing the surviving family members into the rental market. This forced transition would generate new, recurrent housing costs, mechanically exacerbating rather than alleviating the overall liquid toxicity. Consequently, housing wealth represents structurally locked capital, thus failing to serve as a functional last-resort financial shield against institutional voids.
	
	Consequently, the true determinant of EoL economic vulnerability is not the mere absence of accumulated assets, but acute liquidity constraints. This is definitely proven by the massive penalty associated with subjective financial distress. By driving an additional burden of $+4,771.7$ PPS € on average, and escalating to $+6,100.8$ PPS € at the extreme tail, this variable mathematically demonstrates that the collapse of the palliative shield is fundamentally mediated by the household's immediate inability to generate cash flow. Ultimately, this reinforces the structural disconnect between physical assets and liquid capacity: when confronted with acute financial distress, a primary residence cannot be deployed as a functional shield, leaving illiquid households completely defenseless against the massive costs of terminal care.
	
	Finally, this theoretical framework is definitively validated by the COVID-19 systemic stress test. By isolating the pandemic shock through a dedicated fixed effect, we demonstrated that the underlying architecture of socio-institutional inequality remained completely intact. The pandemic did not create new rules; rather, as an exogenous shock that fractured formal domiciliary supply chains \citep{Werner2020}, it acted as a brutal magnifier of pre-existing tail risks. While relatively muted at the median, the COVID-19 penalty exploded at the 75th ($+4,376.1$ PPS €) and 90th ($+10,738.8$ PPS €) percentiles, perfectly capturing the unprecedented price volatility and friction of shadow care markets during global lockdowns.
	
	Finally, this theoretical framework is definitively validated by the COVID-19 systemic stress test. By isolating the pandemic shock, we demonstrated that the underlying architecture of socio-institutional inequality remained completely intact. The pandemic did not create new rules; rather, as an exogenous shock that fractured formal domiciliary supply chains \citep{Werner2020}, it acted as a magnifier of pre-existing tail risks. While relatively muted at the median, the COVID-19 penalty exploded at the 75th ($+4,376.1$ PPS €) and 90th ($+10,738.8$ PPS €) percentiles. Crucially, our decomposition (see \ref{app:heterogeneity}) shows that this right tail shock was entirely driven by time poverty.
	
	The ultimate policy implication is the necessity of conditional expansion. European authorities should dictate the decentralization of EoL care to be strictly conditional upon simultaneous investments in formal LTC infrastructures. Expanding home-based care without establishing robust, state-funded LTC schemes fundamentally acts as a regressive policy that disproportionately penalizes lower-income households \citep{Ilinca2017} and exacerbates broader healthcare pressures \citep{CostaFont2018}.
	
	\subsection{Limitations}
	
	While our SDG framework robustly addresses the fundamental problem of causal inference, this study is subject to several data-driven limitations inherent to the SHARE End-of-Life module. 
	
	First, there is a structural temporal mismatch between our primary outcome and the treatment indicator. While direct OOP expenditures are recorded over the entire last year of life, PC utilization is captured exclusively for the final four weeks. Although modern clinical frameworks advocate for the early integration of PC months or even years before death \citep{Radbruch2020}, the survey design restricts our observation to the strict terminal phase. Thus, while our digital twin architecture evaluates the counterfactual economic trajectory conditioned on this final intervention, this temporal discrepancy may introduce alignment noise and obscure the broader financial benefits of early PC integration. 
	
	Second, the SHARE questionnaire aggregates domiciliary palliative care and institutional hospice care into a single and indivisible indicator. Consequently, our causal estimates capture the aggregate protective effect of community-oriented and hospice-based PC against standard care. This aggregation prevents a granular decomposition of the specific cost-saving mechanisms, making it impossible to disentangle the exact liquid toxicity and labor substitution dynamics between home care and dedicated hospice facilities.
	
	Finally, while our analysis rigorously controls for multiple covariates, the estimates on EoL variables rely on proxy-reported retrospective data. Although proxy interviews are the established gold standard for EoL economic evaluation, they may be subject to recall bias, particularly regarding the exact quantification of intense informal caregiving hours during emotionally distressing terminal phases. Future research leveraging linked administrative claims data could further validate the magnitude of the tail events identified by our quantile models.
	
	\section{Conclusion} 
	
	This study contributes to the health economics of EoL care by providing causal estimates of the socioeconomic impact of palliative care across Europe. Moving beyond traditional observational correlations, we leveraged a novel Synthetic Data Generation framework to simulate a complete counterfactual manifold, thereby challenging the strict cost-shifting hypothesis on a macroscopic level. We demonstrate that, when adequately provisioned, PC has the potential to act as a ``double shield'', successfully truncating the right tail of OOP expenditures while simultaneously mitigating the time poverty associated with intense informal caregiving.
	
	However, this narrative of success is fractured by systemic inequalities and extreme tail events. For the top 10\% of households facing the most grueling terminal trajectories, this protection collapses, driving an average economic penalty exceeding $+39,000$ PPS €. At this extreme margin, the sheer volume of 24/7 care requirements overwhelms the system. Whether households absorb this shock through informal labor to cover domiciliary care voids, or through huge OOP co-payments for hospice facilities, the result is identical: a brutal cost-shifting mechanism where the state's cost-containment is fully subsidized by the family's financial and temporal exhaustion. The protective capacity of the shield is therefore not an inherent property of the medical intervention itself, but is strictly contingent upon the welfare architecture, the clinical trajectory, and the socioeconomic vulnerabilities of the household. Our granular models reveal that the baseline relief is—as expected—highly sensitive to underlying financial distress. We expose a ``broken shield'' reality for patients dying from prolonged, non-cancer trajectories—such as severe dementia or organ failure—where economic penalties explode at the extreme tails of the distribution. While lacking a spousal safety net exposes patients to upfront vulnerabilities, rigid gender norms trigger immediate labor market ejection for male caregivers and hidden, long-term occupational hysteresis for females.
	
	Integrating the unprecedented exogenous shock of the COVID-19 pandemic allowed us to rigorously separate transient market friction from permanent structural failures. The pandemic acted as a systemic stress test, magnifying the OOP expenditures at the extreme tails of the distribution without altering the fundamental architecture of European EoL inequality. At the macro-institutional level, our models expose a fragmented landscape where the Nordic regime imposes significant economic penalties driven by the opportunity cost of withdrawing from a high-wage labor market, the Southern regime forces a massive surge in labor substitution, and Eastern European systems drive the most vulnerable households toward absolute resource exhaustion. 
	
	Ultimately, this paper serves as a warning to European policymakers. Expanding access to palliative care is an ethical imperative, but doing it in institutional vacuums is an incomplete and dangerous policy. To eradicate disease-based inequality, health systems must decouple palliative models from the historical oncological paradigm and aggressively invest in formal domiciliary support networks. Palliative care is a fundamental human right, but without equitable welfare structures to sustain it—and protect against both predictable decline and unforeseen macroeconomic shocks—it risks becoming a privilege that protects the robust while abandoning the most vulnerable to disease-induced poverty.
	
	\appendix
	
	\section*{Contributors}
	All authors conceived and designed the reported study. PG led the data analysis and data augmentation; all authors contributed to the interpretation of results. PG and EP drafted the manuscript; all authors reviewed it critically for important intellectual content. All authors provided final approval of the version to be published. All authors agree to be accountable for all aspects of the work in ensuring that questions related to the accuracy or integrity of any part of the work are appropriately investigated and resolved. PG has directly accessed and verified the underlying data reported in the manuscript. All authors had full access to all of the data in the study and accept responsibility to submit the Article for publication.
	
	\bibliographystyle{unsrtnat}
	\bibliography{EHEW}

@article{Knaul2018,
	title = {Alleviating the access abyss in palliative care and pain relief—an imperative of universal health coverage: the {Lancet Commission} report},
	author = {Knaul, Felicia M and others},
	journal = {The Lancet},
	volume = {391},
	number = {10128},
	pages = {1391--1454},
	year = {2018},
	publisher = {Elsevier}
}

@article{Carrera2018,
	title={The financial burden and distress of patients with cancer: Understanding and stepping-up action on the financial toxicity of cancer treatment},
	author={Carrera, Pricivel M and Kantarjian, Hagop M and Blinder, Victoria S},
	journal={CA: Cancer J Clin},
	volume={68},
	number={2},
	pages={153--165},
	year={2018},
	publisher={Wiley Online Library}
}

@article{Urwin2021,
	title={The monetary valuation of informal care to cancer decedents at end-of-life: Evidence from a national census survey},
	author={Urwin, Sean and others},
	journal={Palliat Med},
	volume={35},
	number={4},
	pages={750--758},
	year={2021},
	publisher={SAGE Publications Sage UK: London, England}
}

@article{May2018,
	title={Economics of Palliative Care for Hospitalized Adults With Serious Illness: A Meta-analysis},
	author={May, Peter and others},
	journal={JAMA Intern Med},
	volume={178},
	number={6},
	pages={820--829},
	year={2018},
	publisher={American Medical Association}
}

@article{Mihaylova2011,
	title={Review of statistical methods for analysing healthcare resources and costs},
	author={Mihaylova, Borislava and Briggs, Andrew and O'Hagan, Anthony and Thompson, Simon G},
	journal={Health Econ},
	volume={20},
	number={8},
	pages={897--916},
	year={2011},
	publisher={Wiley Online Library}
}

@article{Aoun2005,
	title={Evidence in palliative care research: how should it be gathered?},
	author={Aoun, Samar M and Kristjanson, Linda J},
	journal={Med J Aust},
	volume={183},
	number={5},
	pages={264--266},
	year={2005},
	publisher={Wiley}
}

@article{Fassbender2009,
	title={Cost trajectories at the end of life: the Canadian experience},
	author={Fassbender, Konrad and Fainsinger, Robin L and Carson, Mary and Finegan, Barry A},
	journal={J Pain Symptom Manage},
	volume={38},
	number={1},
	pages={75--80},
	year={2009},
	publisher={Elsevier}
}

@article{GomezBatiste2017,
	title={Building integrated palliative care programs and services},
	author={G{\'o}mez-Batiste, Xavier and Connor, Stephen},
	journal={Catalonia Collaborating Centre for Palliative Care},
	year={2017},
	publisher={Liberd{\'u}plex}
}

@article{Gardiner2020,
	title={Equity and the financial costs of informal caregiving in palliative care: a critical debate},
	author={Gardiner, Clare and others},
	journal={BMC Palliat Care},
	volume={19},
	number={1},
	pages={71},
	year={2020},
	publisher={Springer}
}

@article{Gardiner2014,
	title={Exploring the financial impact of caring for family members receiving palliative and end-of-life care: a systematic review of the literature},
	author={Gardiner, Clare and others},
	journal={Palliat Med},
	volume={28},
	number={3},
	pages={375--390},
	year={2014},
	publisher={SAGE Publications Sage UK: London, England}
}

@article{Radbruch2020,
	title={Redefining palliative care—A new consensus-based definition},
	author={Radbruch, Lukas and others},
	journal={J Pain Symptom Manage},
	volume={60},
	number={4},
	pages={754--764},
	year={2020},
	publisher={Elsevier}
}

@article{Sleeman2019,
	title={The escalating global burden of serious health-related suffering: projections to 2060 by world regions, age groups, and health conditions},
	author={Sleeman, Katherine E and others},
	journal={Lancet Glob Health},
	volume={7},
	number={7},
	pages={e883--e892},
	year={2019},
	publisher={Elsevier}
}

@article{Bambra2011,
	title = {Health inequalities and welfare state regimes: theoretical insights on a public health `puzzle'},
	author = {Bambra, Clare},
	journal = {J Epidemiol Community Health},
	volume = {65},
	number = {9},
	pages = {740--745},
	year = {2011},
	publisher = {BMJ Publishing Group Ltd}
}

@article{Husereau2022,
	title = {Consolidated Health Economic Evaluation Reporting Standards 2022 (CHEERS 2022) statement: updated reporting guidance for health economic evaluations},
	author = {Husereau, Don and others},
	journal = {Value Health},
	volume = {25},
	number = {1},
	pages = {3--9},
	year = {2022}
}

@article{Hayman2001,
	title = {Estimating the cost of informal caregiving for elderly patients with cancer},
	author = {Haynman, James A and others},
	journal = {J Clin Oncol},
	volume = {19},
	number = {13},
	pages = {3219--3225},
	year = {2021}
}

@article{Boogaard2019,
	title = {How Is End-of-Life Care With and Without Dementia Associated With Informal Caregivers' Outcomes?},
	author = {Husereau, Don and others},
	journal = {Am J Osp Palliat Care},
	volume = {36},
	number = {11},
	pages = {1008--1015},
	year = {2019}
}

@article{Fischer2024,
	title={Methodological challenges and potential solutions for economic evaluations of palliative and end-of-life care: A systematic review},
	author={Fischer, Claudia and Bednarz, Damian and Simon, Judit},
	journal={Palliat Med},
	volume={38},
	number={1},
	pages={85--99},
	year={2024},
	publisher={SAGE Publications Sage UK: London, England}
}

@article{AtheyImbens2017,
	title={The state of applied econometrics: Causality and policy evaluation},
	author={Athey, Susan and Imbens, Guido W},
	journal={Journal of Economic Perspectives},
	volume={31},
	number={2},
	pages={3--32},
	year={2017},
	publisher={American Economic Association}
}

@inproceedings{Kotelnikov2023,
	title={TabDDPM: Modelling Tabular Data with Diffusion Models},
	author={Kotelnikov, Akim and Baranchuk, Dmitry and Rubachev, Ivan and Babenko, Artem},
	booktitle={International Conference on Machine Learning},
	volume = {40},
	pages={17564--17579},
	year={2023},
	organization={PMLR}
}

@article{Thevenon2011,
	title={Family policies in {OECD} countries: A comparative analysis},
	author={Th{\'e}venon, Olivier},
	journal={Population and Development Review},
	volume={37},
	number={1},
	pages={57--87},
	year={2011},
	publisher={Wiley Online Library}
}

@article{Korpi2013,
	title={Women's opportunities under different family policy constellations: Gender, class, and inequality tradeoffs in western countries re-examined},
	author={Korpi, Walter and Ferrarini, Tommy and Englund, Stefan},
	journal={Social Politics: International Studies in Gender, State \& Society},
	volume={20},
	number={1},
	pages={1--40},
	year={2013},
	publisher={Oxford University Press}
}

@article{Smith2014,
	title={Evidence on the cost and cost-effectiveness of palliative care: a literature review},
	author={Smith, Samantha and Brick, Aoife and O'Hara, Sinéad and Normand, Charles},
	journal={Palliative Medicine},
	volume={28},
	number={2},
	pages={130--150},
	year={2014},
	publisher={SAGE Publications Sage UK: London, England}
}

@article{VanDenBerg2004,
	title={Economic valuation of informal care. An overview of methods and applications},
	author={Van den Berg, Bernard and Brouwer, Werner and Koopmanschap, Marc},
	journal={Eur J Health Econ},
	volume={5},
	number={1},
	pages={36--45},
	year={2004},
	publisher={Springer}
}

@article{Cassel2018,
	title={Effect of a Home-Based Palliative Care Program on Healthcare Use and Costs},
	author={Cassel, J Brian and others},
	journal={J Am Geriatr Soc},
	volume={64},
	number={11},
	pages={2288--2295},
	year={2016},
	publisher={Wiley Online Library}
}

@book{EspingAndersen1990,
	title={The Three Worlds of Welfare Capitalism},
	author={Esping-Andersen, G{\o}sta},
	year={1990},
	publisher={Princeton University Press},
	address={Princeton, NJ}
}

@article{Ferrera1996,
	title={The ‘Southern model’ of welfare in social Europe},
	author={Ferrera, Maurizio},
	journal={Journal of European Social Policy},
	volume={6},
	number={1},
	pages={17--37},
	year={1996},
	publisher={Sage Publications Sage CA: Thousand Oaks, CA}
}

@book{Pearl2009,
	title={Causality: Models, Reasoning, and Inference},
	author={Pearl, Judea},
	year={2009},
	edition={2nd},
	publisher={Cambridge University Press},
	address={Cambridge, UK}
}

@article{Chernozhukov2018,
	title={Double/debiased machine learning for treatment and structural parameters},
	author={Chernozhukov, Victor and others},
	journal={The Econometrics Journal},
	volume={21},
	number={1},
	pages={C1--C68},
	year={2018},
	publisher={Oxford University Press}
}

@article{Rechel2011,
	title={The Soviet legacy in diagnosis and treatment: implications for population health},
	author={Rechel, Boika and Kennedy, Colin and McKee, Martin and Rechel, Bernd},
	journal={Journal of Public Health Policy},
	volume={32},
	number={3},
	pages={293--304},
	year={2011},
	publisher={Springer}
}

@inproceedings{niu2025fest,
	title={FEST: A Unified Framework for Evaluating Synthetic Tabular Data},
	author={Niu, Weijie and Huertas Celdran, Alberto and Siarsky, Karoline and Stiller, Burkhard},
	booktitle={Proceedings of the International Conference on Information Systems Security and Privacy},
	pages={434--444},
	year={2025}
}

@article{Wagstaff2018,
	title={Progress on catastrophic health spending in 133 countries: a retrospective observational study},
	author={Wagstaff, Adam and others},
	journal={Lancet Glob Health},
	volume={6},
	number={2},
	pages={e169--e179},
	year={2018},
	publisher={Elsevier}
}

@article{Etkind2017,
	title={How many people will need palliative care in 2040? Past trends, future projections and implications for services},
	author={Etkind, Simon Noah and others},
	journal={BMC Medicine},
	volume={15},
	number={102},
	year={2017},
	publisher={Springer}
}

@article{MadleyDowd2019,
	title={The proportion of missing data should not be used to guide decisions on multiple imputation},
	author={Madley-Dowd, Paul and Hughes, Rachael and Tilling, Kate and Heron, Jon},
	journal={J Clin Epidemiol},
	volume={110},
	pages={63--73},
	year={2019},
	publisher={Elsevier}
}

@article{KingNielsen2019,
	title={Why propensity scores should not be used for matching},
	author={King, Gary and Nielsen, Richard},
	journal={Political Analysis},
	volume={27},
	number={4},
	pages={435--454},
	year={2019},
	publisher={Cambridge University Press}
}

@book{ImbensRubin2015,
	title={Causal inference in statistics, social, and biomedical sciences},
	author={Imbens, Guido W and Rubin, Donald B},
	year={2015},
	publisher={Cambridge University Press}
}

@article{Kunzel2019,
	title={Metalearners for estimating heterogeneous treatment effects using machine learning},
	author={K{\"u}nzel, S{\"o}ren R and Sekhon, Jasjeet S and Bickel, Peter J and Yu, Bin},
	journal={Proc Natl Acad Sci USA},
	volume={116},
	number={10},
	pages={4156--4165},
	year={2019},
	publisher={National Academy of Sciences}
}

@article{BorschSupan2013,
	title={Data resource profile: the Survey of Health, Ageing and Retirement in Europe (SHARE)},
	author={B{\"o}rsch-Supan, Axel and Brandt, Martina and Hunkler, Christian and Kneip, Thorsten and Korbmacher, Julie and Malter, Frederic and Schaan, Barbara and Stuck, Stephanie and Zuber, Sabrina},
	journal={International Journal of Epidemiology},
	volume={42},
	number={4},
	pages={992--1001},
	year={2013},
	publisher={Oxford University Press}
}

@article{Floridi2022,
	title={Partner care arrangements and well-being in mid-and later life: The role of gender across care contexts},
	author={Floridi, Ginevra and Quashie, Nekehia T and Glaser, Karen and Brandt, Martina},
	journal={J Gerontol B Psychol Sci Soc Sci},
	volume={77},
	number={2},
	pages={435--445},
	year={2022},
	publisher={Oxford University Press}
}

@article{Vignoli2025,
	title={Partners' health and silver splits in Europe: A gendered pattern?},
	author={Vignoli, Daniele and Alderotti, Giammarco and Tomassini, Cecilia},
	journal={Journal of Marriage and Family},
	volume={87},
	number={4},
	pages={1639--1663},
	year={2025},
	publisher={Wiley Online Library}
}

@article{Poterba2017,
	title={The asset cost of poor health},
	author={Poterba, James M and Venti, Steven F and Wise, David A},
	journal={The Journal of the Economics of Ageing},
	volume={9},
	pages={172--184},
	year={2017},
	publisher={Elsevier}
}

@article{Albertini2025,
	title={Caring in the XXI century: the sustainability of long-term care in aging societies-mapping challenges and developing solutions within the Age-It Research Program},
	author={Albertini, Marco and others},
	journal={J Gerontol B Psychol Sci Soc Sci},
	year={2025},
	publisher={Oxford University Press}
}

@article{ManningMullahy2001,
	title={Estimating log models: to transform or not to transform?},
	author={Manning, Willard G and Mullahy, John},
	journal={J Health Econ},
	volume={20},
	number={4},
	pages={461--494},
	year={2001},
	publisher={Elsevier}
}

@article{Murray2005,
	title={Illness trajectories and palliative care},
	author={Murray, Scott A and Kendall, Marilyn and Boyd, Kirsty and Sheikh, Aziz},
	journal={BMJ},
	volume={330},
	number={7498},
	pages={1007--1011},
	year={2005},
	publisher={British Medical Journal Publishing Group}
}

@article{Cinelli2020,
	title={Making Sense of Sensitivity: Extending Omitted Variable Bias},
	author={Cinelli, Carlos and Hazlett, Chad},
	journal={JRSS: Series B (Statistical Methodology)},
	volume={82},
	number={1},
	pages={39--67},
	year={2020},
	publisher={Wiley Online Library}
}

@article{Holland1986,
	title={Statistics and Causal Inference},
	author={Holland, Paul W},
	journal={JASA},
	volume={81},
	number={396},
	pages={945--960},
	year={1986},
	publisher={Taylor \& Francis}
}

@article{Bergmann2021,
	title={The Impact of COVID-19 on Informal Caregiving and Care Receiving Across Europe During the First Phase of the Pandemic},
	author={Bergmann, Michael and Wagner, Melanie},
	journal={Front Public Health},
	volume={9},
	pages={673874},
	year={2021},
	publisher={Lausanne: Frontiers Editorial Office}
}

@unpublished{Grassi2026,
	author = {Grassi, Pietro and Bellini, Arianna and Seghieri, Chiara and Vignoli, Daniele},
	title = {Inequalities in Access to Palliative Care During the Final Month of Life in Europe: A Multilevel Analysis},
	year = {2026},
	note = {Mimeo}
}

@article{VanBuuren2011,
	title={mice: Multivariate imputation by chained equations in R},
	author={van Buuren, Stef and Groothuis-Oudshoorn, Catharina GM},
	journal={Journal of Statistical Software},
	volume={45},
	number={3},
	year={2011},
	publisher={University of California at Los Angeles}
}

@article{Yale2020,
	title={Generation and evaluation of privacy preserving synthetic health data},
	author={Yale, Andrew and others},
	journal={Neurocomputing},
	volume={416},
	pages={244--255},
	year={2020},
	publisher={Elsevier}
}

@misc{Eurostat2021,
	title={Purchasing power parities, price level indices, nominal and real expenditures by analytical categories - based on COICOP 2018},
	author={{Eurostat}},
	year={2026},
	howpublished={\url{https://ec.europa.eu/eurostat/databrowser/view/prc_ppp_ind_1}},
	note={Accessed: 2026-03-17}
}

@article{Efron1987,
	title={Better bootstrap confidence intervals},
	author={Efron, Bradley},
	journal={Journal of the American Statistical Association},
	volume={82},
	number={397},
	pages={171--185},
	year={1987},
	publisher={Taylor \& Francis}
}

@article{CameronMiller2015,
	title={A practitioner's guide to cluster-robust inference},
	author={Cameron, A Colin and Miller, Douglas L},
	journal={Journal of Human Resources},
	volume={50},
	number={2},
	pages={317--372},
	year={2015},
	publisher={University of Wisconsin Press}
}

@article{BellMcCaffrey2002,
	title={Bias reduction in standard errors for linear regression with multi-stage samples},
	author={Bell, Robert M and McCaffrey, Daniel F},
	journal={Survey Methodology},
	volume={28},
	number={2},
	pages={169--181},
	year={2002}
}

@article{Koenker1978,
	title={Regression quantiles},
	author={Koenker, Roger and Bassett Jr, Gilbert},
	journal={Econometrica},
	volume={46},
	number={1},
	pages={33--50},
	year={1978},
	publisher={JSTOR}
}

@article{Machado2005,
	title={Quantiles for counts},
	author={Machado, Jos{\'e} AF and Santos Silva, JMC},
	journal={Journal of the American Statistical Association},
	volume={100},
	number={472},
	pages={1226--1237},
	year={2005},
	publisher={Taylor \& Francis}
}

@article{OlivaMoreno2019,
	title={The economic value of time of informal care and its determinants (The CUIDARSE Study)},
	author={Oliva-Moreno, Juan and others},
	journal={PLoS One},
	volume={14},
	number={5},
	pages={e0217016},
	year={2019},
	publisher={PLOS}
}

@misc{Who2020,
	title = {Palliative Care},
	author = {{World Health Organization}},
	year = {2020},
	howpublished = {\url{https://www.who.int/news-room/fact-sheets/detail/palliative-care}},
	note = {Accessed: 2026–03–19}
}

@article{Joo2017,
	title={Economic Burden of Informal Caregiving Associated With History of Stroke and Falls Among Older Adults in the U.S.},
	author={Joo, Heesoo and others},
	journal={AJPM},
	volume={53},
	number={6},
	pages={S197-S204},
	year={2017},
	publisher={Elsevier}
}

@article{Koopmanschap2008,
	title={An overview of methods and applications to value informal care in economic evaluations of healthcare},
	author={Koopmanschap, Marc A and van Exel, Job N and van den Berg, Bernard and Brouwer, Werner BF},
	journal={Pharmacoeconomics},
	volume={26},
	number={4},
	pages={269--280},
	year={2008},
	publisher={Springer}
}

@article{OlivaMoreno2017,
	title={The Valuation of Informal Care in Cost-of-Illness Studies: A Systematic Review},
	author={Oliva-Moreno, Juan and Trapero-Bertran, Marta and Pe{\~n}a-Longobardo, Luz Maria and Del Pozo-Rubio, Ra{\'u}l},
	journal={PharmacoEconomics},
	volume={35},
	number={3},
	pages={331--345},
	year={2017},
	publisher={Springer}
}

@article{Heger2020,
	title={Short- and Medium-Term Effects of Informal Eldercare on Labor Market Outcomes},
	author={Heger, D{\"o}rte and Korfhage, Thorben},
	journal={Feminist Economics},
	volume={26},
	number={4},
	pages={205--227},
	year={2020},
	publisher={Routledge}
}

@article{CostaFont2015,
	title={Financing long-term care: ex-ante, ex-post or both?},
	author={Costa-Font, Joan and Courbage, Christophe and Swartz, Katherine},
	journal={J Health Econ},
	volume={24},
	number={S1},
	pages={45--57},
	year={2015},
	publisher={Wiley Online Library}
}

@article{Bonsang2008,
	title={Does informal care from children to their elderly parents substitute for formal care in Europe?},
	author={Bonsang, Eric},
	journal={J Health Econ},
	volume={28},
	number={1},
	pages={143--154},
	year={2009},
	publisher={Elsevier}
}

@article{Arrow1963,
	title={Uncertainty and the Welfare Economics of Medical Care},
	author={Arrow, Kenneth J},
	journal={The American Economic Review},
	volume={53},
	number={5},
	pages={941--973},
	year={1963},
	publisher={American Economic Association}
}

@article{Blau2013,
	title={Female Labor Supply: Why Is the United States Falling Behind?},
	author={Blau, Francine D and Kahn, Lawrence M},
	journal={The American Economic Review},
	volume={103},
	number={3},
	pages={251--256},
	year={2013},
	publisher={American Economic Association}
}

@article{Lott2016,
	title={Gender Discrepancies in the Outcomes of Schedule Control on Overtime Hours and Income in Germany},
	author={Lott, Yvonne and Chung, Heejung},
	journal={European Sociological Review},
	volume={32},
	number={6},
	pages={752--765},
	year={2016},
	publisher={Oxford University Press}
}

@article{Chung2020,
	title={Flexible Working, Work-Life Balance, and Gender Equality: Introduction},
	author={Chung, Heejung and van der Lippe, Tanja},
	journal={Soc Indic Res},
	volume={151},
	number={2},
	pages={365--381},
	year={2020},
	publisher={Springer}
}

@article{Canta2012,
	title={Long term care insurance and family norms},
	author={Canta, Chiara and Pestieau, Pierre},
	journal={The B.E. Journal of Economic Analysis \& Policy},
	volume={14},
	number={2},
	year={2013},
	pages={401--428},
	publisher={De Gruyter}
}

@techreport{NICE2025,
	title={NICE technology appraisal and highly specialised technologies guidance: the
	manual},
	author={{National Institute for Health and Care Excellence}},
	institution={NICE},
	number={PMG36},
	year={2025},
	howpublished={\url{https://www.nice.org.uk/process/pmg36}},
	note = {Accessed: 2026–03–24}
}

@article{Szebehely2018,
	title={Nordic eldercare—weak universalism becoming weaker?},
	author={Szebehely, Marta and Meagher, Gabrielle},
	journal={Journal of European Social Policy},
	volume={28},
	number={3},
	pages={294--308},
	year={2018},
	publisher={SAGE Publications Sage UK: London, England}
}

@article{Fadlon2019,
	title={Family health behaviors},
	author={Fadlon, Itzik and Nielsen, Torben Heien},
	journal={American Economic Review},
	volume={109},
	number={9},
	pages={3162--3191},
	year={2019},
	publisher={American Economic Association}
}

@article{DeNardi2016,
	title={Medicaid insurance in old age},
	author={De Nardi, Mariacristina and French, Eric and Jones, John Bailey},
	journal={American Economic Review},
	volume={106},
	number={11},
	pages={3480--3520},
	year={2016},
	publisher={American Economic Association}
}

@article{Ilinca2017,
	title={Fairness and Eligibility to Long-Term Care: An Analysis of the Factors Driving Inequality and Inequity in the Use of Home Care for Older Europeans},
	author={Ilinca, Stefania and Rodrigues, Ricardo and Schmidt, Andrea E},
	journal={Int J Environ Res Public Health},
	volume={14},
	number={10},
	pages={1224},
	year={2017},
	publisher={MDPI}
}

@article{CostaFont2018,
	title={Does long-term care subsidization reduce hospital admissions and utilization?},
	author={Costa-Font, Joan and Jimenez-Martin, Sergi and Vilaplana, Cristina},
	journal={J Health Econ},
	volume={58},
	pages={43--66},
	year={2018},
	publisher={Elsevier}
}

@article{Werner2020,
	title={Long-Term Care Policy after Covid-19 — Solving the Nursing Home Crisis},
	author={Werner, Rachel M and Hoffman, Allison K and Coe, Norma B},
	journal = {New England Journal of Medicine},
	volume = {383},
	number = {10},
	pages = {903-905},
	year = {2020},
	publisher={Mass Medical Soc}
}

@article{Bambra2020,
	title={The COVID-19 pandemic and health inequalities},
	author={Bambra, Clare and Riordan, Ryan and Ford, John and Matthews, Fiona},
	journal={J Epidemiol Community Health},
	volume={74},
	number={11},
	pages={964--968},
	year={2020},
	publisher={BMJ Publishing Group Ltd}
}

@article{Grabowski2020,
	title={Nursing Home Care in Crisis in the Wake of COVID-19},
	author={Grabowski, David C and Mor, Vincent},
	journal={JAMA},
	volume={324},
	number={1},
	pages={23--24},
	year={2020},
	publisher={American Medical Association}
}

@article{Bekelman2016,
	title={Comparison of Site of Death, Health Care Utilization, and Hospital Expenditures for Patients Dying With Cancer in 7 Developed Countries},
	author={Bekelman, Justin E and others},
	journal={JAMA},
	volume={315},
	number={3},
	pages={272--283},
	year={2016},
	publisher={American Medical Association}
}

@article{French2017,
	title={End-Of-Life Medical Spending In Last Twelve Months Of Life Is Lower Than Previously Reported},
	author={French, Eric B and others},
	journal={Health Affairs},
	volume={36},
	number={7},
	pages={1211--1217},
	year={2017},
	publisher={Project HOPE}
}

@article{Hoefman2013,
	title={How to Include Informal Care in Economic Evaluations},
	author={Hoefman, Renske J and van Exel, Job and Brouwer, Werner},
	journal={Pharmacoeconomics},
	volume={31},
	number={12},
	pages={1105--1119},
	year={2013},
	publisher={Springer}
}

@article{Grundy2001,
	title={The socioeconomic status of older adults: How should we measure it in studies of health inequalities?},
	author={Grundy, Emily and Holt, G},
	journal={Journal of Epidemiology and Community Health},
	volume={55},
	number={12},
	pages={895--904},
	year={2001},
	publisher={BMJ Publishing Group}
}
	
	\section*{Code Availability Statement}
	The \texttt{R} code used for the data cleaning, multiple imputation, and statistical modeling is openly available at \url{https://github.com/pietrograssi-unifi/EHEW2026}.
	
	\newpage
	
	\section*{SHARE acknowledgements}
	This paper uses data from SHARE Waves 1, 2, 3, 4, 5, 6, 7, 8 and 9  (DOIs:  \url{10.6103/SHARE.w1.900}, \url{10.6103/SHARE.w2.900}, \url{10.6103/SHARE.w3.900}, \url{10.6103/SHARE.w4.900}, \url{10.6103/SHARE.w5.900}, \url{10.6103/SHARE.w6.900}, \url{10.6103/SHARE.w6.DBS.100}, \url{10.6103/SHARE.w7.900}, \url{10.6103/SHARE.w8.900}, \url{10.6103/SHARE.w8ca.900}, \url{10.6103/SHARE.w9.900}, \url{10.6103/SHARE.w9ca900}, \url{10.6103/SHARE.HCAP1.100}) see \citet{BorschSupan2013} for methodological details.
	
	The SHARE data collection has been funded by the European Commission, DG RTD through FP5 (QLK6-CT-2001-00360), FP6 (SHARE-I3: RII-CT-2006-062193, COMPARE: CIT5-CT-2005-028857, SHARELIFE: CIT4-CT-2006-028812), FP7 (SHARE-PREP: GA N°211909, SHARE-LEAP: GA N°227822, SHARE M4: GA N°261982, DASISH: GA N°283646) and Horizon 2020 (SHARE-DEV3: GA N°676536, SHARE-COHESION: GA N°870628, SERISS: GA N°654221, SSHOC: GA N°823782, SHARE-COVID19: GA N°101015924) and by DG Employment, Social Affairs \& Inclusion through VS 2015/0195, VS 2016/0135, VS 2018/0285, VS 2019/0332, VS 2020/0313, SHARE-EUCOV: GA N°101052589 and EUCOVII: GA N°101102412. Additional funding from the German Federal Ministry of Research, Technology and Space (01UW1301, 01UW1801, 01UW2202), the Max Planck Society for the Advancement of Science, the U.S. National Institute on Aging (U01\_AG09740-13S2, P01\_AG005842, P01\_AG08291, P30\_AG12815, R21\_AG025169, Y1-AG-4553-01, IAG\_BSR06-11, OGHA\_04-064, BSR12-04, R01\_AG052527-02, R01\_AG056329-02, R01\_AG063944, HHSN271201300071C, RAG052527A) and from various national funding sources is gratefully acknowledged (see \url{www.share-eric.eu}).
	
	This paper uses data from the generated easySHARE data set (DOI: \url{10.6103/SHARE.easy.900}), see Gruber et al. (2014) for methodological details. The easySHARE release 9.0.0 is based on SHARE Waves 1, 2, 3, 4, 5, 6, 7, 8 and 9 (DOIs: \url{10.6103/SHARE.w1.900}, \url{10.6103/SHARE.w2.900}, \url{10.6103/SHARE.w3.900}, \url{10.6103/SHARE.w4.900}, \url{10.6103/SHARE.w5.900}, \url{10.6103/SHARE.w6.900}, \url{10.6103/SHARE.w7.900}, \url{10.6103/SHARE.w8.900}, \url{10.6103/SHARE.w9.900}).
	
	\newpage
	
	\setcounter{table}{0}
	\setcounter{figure}{0}
	\renewcommand{\thetable}{A\arabic{table}}
	\renewcommand{\thefigure}{A\arabic{figure}}
	
	\section{Sensitivity Analysis on Shadow Pricing}
	\label{app:SHADOW}
	
	A persistent methodological challenge in health economics is the monetization of informal caregiving. While the Proxy Good Method (PGM)—which values unpaid care at the market wage of an equivalent formal substitute—is widely endorsed \citep{VanDenBerg2004, OlivaMoreno2019}, the exact choice of the shadow wage remains subject to debate. Methodologically, the monetization of informal care via the proxy good method (e.g., utilizing average NACE Rev. 2 sector wages) presents inherent structural bounds. Rather than overestimating the macroeconomic burden, this uniform pricing mechanism often acts as a conservative baseline. From a market replacement perspective, average sector wages typically underestimate the true cost of formalizing care, as they exclude employer taxes, benefits, and agency friction costs \citep{Joo2017}. Simultaneously, from a welfare perspective, this approach fails to capture the true opportunity cost for households where caregivers forgo highly lucrative professional careers, thereby structurally flattening the extreme liquid toxicity absorbed by high-income earners \citep{VanDenBerg2004}.
	
	To ensure that our primary finding—the ``double shield'' effect of palliative care on the net total economic burden—is not an artifact of the chosen monetization heuristic, we conducted the following sensitivity analysis. We systematically scaled the country-specific hourly market wage ($W_c$) by a multiplier $m \in \{0.50, 0.75, 1.00, 1.25, 1.50\}$. 
	
	The adjusted Net Average Treatment Effect (NATE) for each multiplier $m$ was calculated across the entire European counterfactual cohort ($N$) as follows:
	$$\text{NATE}(m)=\frac{1}{N}\sum_{i=1}^{N}\left(\Delta OOP_i+\frac{\Delta H_i \times(W_c \times m)}{PPP_c}\right)$$
	where $\Delta OOP_i$ and $\Delta H_i$ represent the estimated Individual Treatment Effects (ITE) for OOP costs and caregiving hours, respectively, and $PPP_c$ represents the dynamic PPP index for country $c$.
	
\begin{figure}[htbp]
		\centering
		\includegraphics[width=\columnwidth]{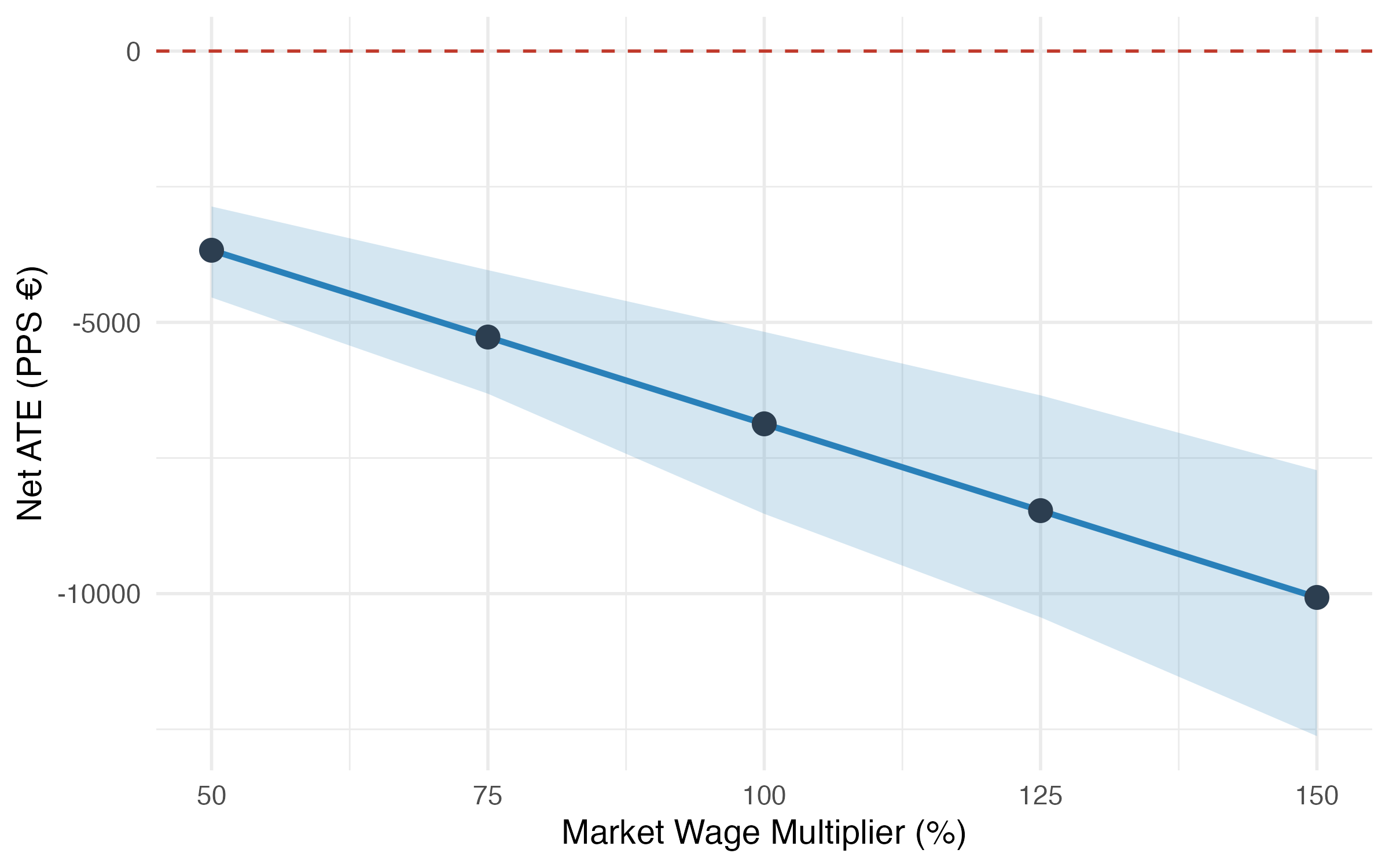}
		\caption{Sensitivity Analysis of the Net Average Treatment Effect to variations in the shadow wage of informal care. The dashed red line indicates the zero-effect threshold. Confidence bounds represent 95\% CI.}
		\label{fig:wage_sensitivity}
	\end{figure}

	The results of the sensitivity estimations, alongside their 95\% confidence intervals, are visualized in Figure \ref{fig:wage_sensitivity}. 
	
	The analysis demonstrates the extreme robustness of the causal estimates. Even when the shadow wage is aggressively depreciated to 50\% of the market rate—effectively simulating a conservative lower-bound scenario where informal care is valued near the absolute minimum wage—the NATE remains solidly and statistically significantly below the zero threshold. Conversely, inflating the wage to 150\% to simulate high opportunity costs logically magnifies the protective economic effect of palliative care.
	
	This confirms that the comprehensive relief provided by palliative care is a structural socio-economic phenomenon. The macroeconomic savings generated by the intervention are not mathematically dependent on the exact monetary valuation of unremunerated domestic labor, effectively shielding the study's core policy implications from controversies surrounding shadow pricing methodologies \citep{Koopmanschap2008, OlivaMoreno2017}.
	
	\section{Sensitivity to Unobserved Confounding}
	\label{app:Cinelli}
	
	A persistent critique of causal models derived from observational survey data is the potential presence of unobserved confounders that might drive the observed economic penalties \citep{AtheyImbens2017}. To rigorously test the resilience of the massive clinical penalty associated with non-cancer terminal trajectories (e.g., severe frailty and dementia), we deployed the state-of-the-art \texttt{sensemakr} analytical suite \citep{Cinelli2020}.
	
	Instead of merely assuming that our rich covariate matrix captures all confounding, we hypothesized the existence of an unobserved ``phantom'' variable. To conduct the most conservative stress test possible, we benchmarked the potential strength of this unknown variable against the observed EoL ADL score (\texttt{eol\_adl\_score}). As demonstrated in our conditional models, severe physical dependency is arguably the most massive mechanical driver of informal care hours and liquid toxicity. 
	
\begin{figure}[htbp]
		\centering
		\includegraphics[width=\columnwidth]{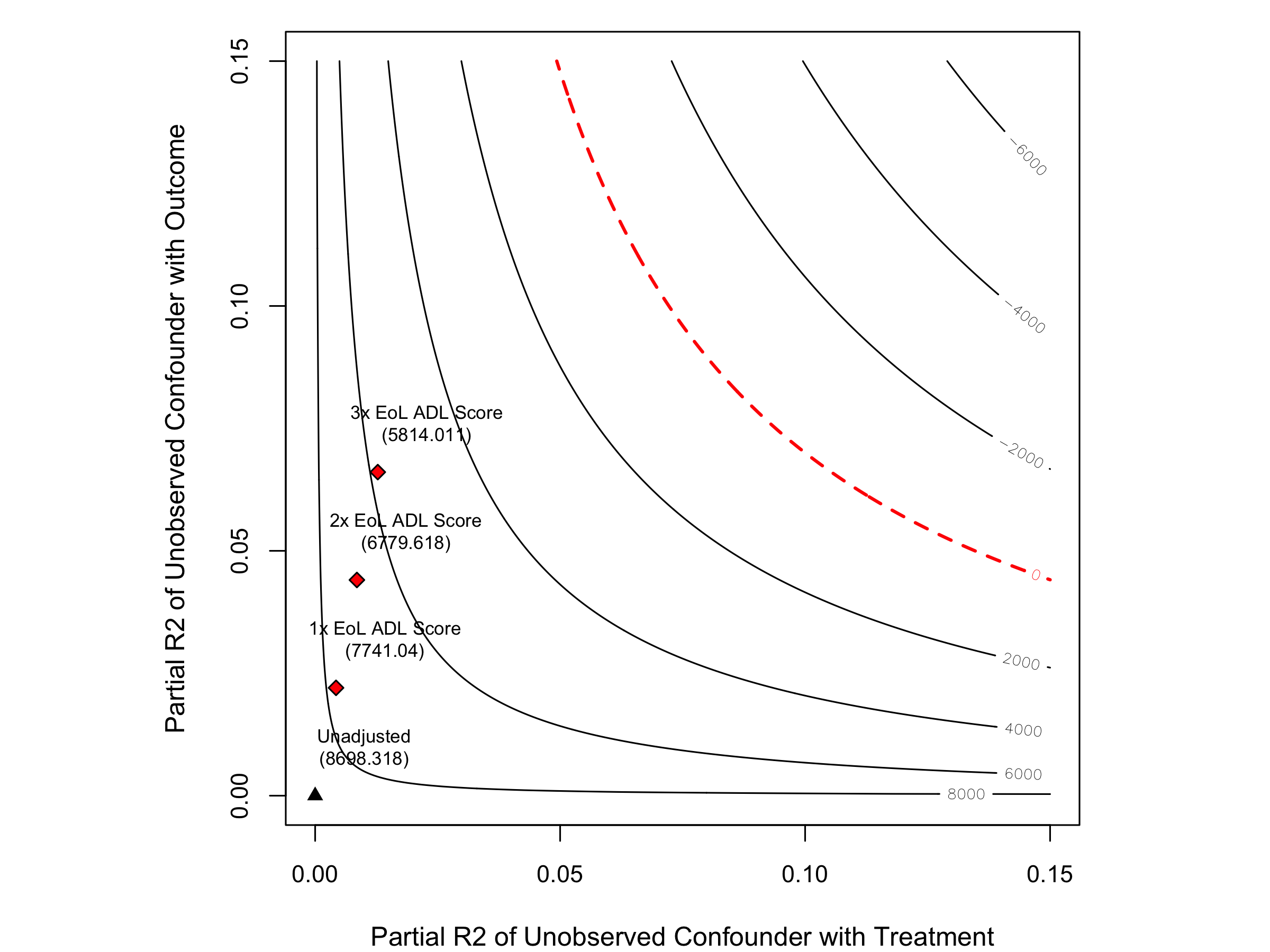}
		\caption{Sensitivity to Unobserved Confounders (Non-Cancer Penalty on Net Burden). The contour plot displays the adjusted point estimate (red lines) and t-value (contour shading) under various strengths of unobserved confounding. The red markers indicate the theoretical bounds if an unobserved confounder were 1x, 2x, or 3x as strong as the observed End-of-Life ADL score.}
		\label{fig:unobserved_confounding}
	\end{figure}

	As illustrated in the contour plot (Figure \ref{fig:unobserved_confounding}), an unobserved confounder would need to explain more than 8.44\% of the residual variance in both the clinical trajectory and the net economic burden to completely eliminate the non-cancer penalty (bringing the point estimate down to zero). Furthermore, it would need to explain at least 4.51\% of the residual variance to render the underlying effect statistically non-significant at the 5\% level. To put this rigorous threshold into perspective, even an unobserved socio-cultural or genetic shock three times as powerful as the patient's entire functional deterioration ($3\times$ \texttt{eol\_adl\_score}) yields an adjusted t-statistic of 2.83, fundamentally failing to drive the underlying statistical signal to zero. This demonstrates that the massive magnitude of the clinical penalty observed across Europe for prolonged dwindling trajectories is a genuine structural dynamic, effectively immune to standard omitted variable bias.
	
	\section{Decoupling Liquid Toxicity from Time Poverty at the Tail}
	\label{app:heterogeneity}
	
	The total net economic burden investigated in our QTE model represents a composite indicator. While evaluating this burden is essential to test the overarching cost-shifting hypothesis, it is useful to decompose the Marginal QTE models in order to observe how specific socio-clinical penalties manifest across distinct economic channels. Table \ref{tab:qte_het} disaggregates the QTE model into the two sides of liquid toxicity and time poverty. The clinical penalty for \texttt{Disease: Other} explodes primarily through the time poverty channel, confirming that prolonged dwindling trajectories mechanically force families to provide impossible volumes of unpaid 24/7 care.
	
	Simultaneously, this decomposition provides empirical insights into how functional dependency, measured by the EoL ADL Score, distinctly shapes the economic burden. At the 90th percentile, the time poverty penalty solidifies into a structural cost of $+2,018.0$ PPS €. This tail dynamic empirically validates the theoretical assertion that serious physical deterioration is a primary driver of time poverty, forcing households into continuous, unremunerated caregiving that cannot be effectively offset by episodic OOP spending on formal domiciliary services.
	
	Furthermore, the QTE decomposition highlights the dynamics of the COVID-19 exogenous shock. While the average CATE indicates that the pandemic acted primarily as an inflationary driver of OOP spending, the QTE models reveal that at the 90th percentile the pandemic violently drove a surge in time poverty. This mirrors the reality of global lockdowns: for the most severe clinical cases, the collapse of formal institutional supply chains could not simply be bypassed by purchasing private care, but forced the household into a default mechanism of massive labor substitution. Finally, the dimensional split validates our hypothesis regarding resource exhaustion in Eastern European regimes: at the 90th percentile, their OOP spending turns highly negative, confirming a hard budget constraint where vulnerable families completely run out of financial liquidity to sustain terminal care.
	
	\begin{table*}[htbp]
		\centering
		\caption{Dimensional Heterogeneity of the QTE on Liquid Toxicity and Time Poverty}
		\label{tab:qte_het}
		\resizebox{0.95\textwidth}{!}{%
			\begin{tabular}{lcccccc}
				\toprule
				& \multicolumn{3}{c}{\textbf{Liquid Toxicity ($\Delta$ OOP Costs)}} & \multicolumn{3}{c}{\textbf{Time Poverty ($\Delta$ Shadow Value)}} \\
				\cmidrule(lr){2-4} \cmidrule(lr){5-7}
				\textbf{Variable} & \textbf{50th} & \textbf{75th} & \textbf{90th} & \textbf{50th} & \textbf{75th} & \textbf{90th} \\
				\midrule
				Intercept & $-1630.273^{***}$ & $190.103$ & $5166.331^{*}$ & $-4099.895$ & $8105.393$ & $31997.730^{*}$ \\
				CoD: Organ Failure & $-9.741$ & $124.392^{+}$ & $329.942$ & $1626.569$ & $3678.656^{*}$ & $1918.571$ \\
				CoD: Other & $-94.121$ & $271.466^{***}$ & $619.636^{*}$ & $6905.228^{***}$ & $13750.243^{***}$ & $17854.139^{***}$ \\
				Welfare: Eastern & $782.616^{***}$ & $-1123.438^{*}$ & $-6775.643^{***}$ & $6566.131^{**}$ & $2861.912$ & $-6875.037$ \\
				Welfare: Nordic & $465.363^{**}$ & $-1574.402^{***}$ & $-7343.004^{***}$ & $2012.459$ & $1868.175$ & $6000.514$ \\
				Welfare: Southern & $-33.901$ & $-1025.333^{*}$ & $-4121.451^{*}$ & $4678.916^{*}$ & $1159.889$ & $-7524.563$ \\
				Wealth: Q2 & $-71.971$ & $-152.868$ & $-344.210$ & $-49.506$ & $-3212.848$ & $-4453.265$ \\
				Wealth: Q3 & $-183.827^{*}$ & $-575.533^{***}$ & $-1025.782^{*}$ & $-659.746$ & $-2332.538$ & $-5493.303$ \\
				Wealth: Q4 (Richest) & $-358.296^{***}$ & $-553.394^{***}$ & $-784.300^{+}$ & $3463.119$ & $179.732$ & $-193.035$ \\
				Age & $12.995^{***}$ & $17.147^{***}$ & $29.730^{**}$ & $-112.416^{+}$ & $-143.742$ & $-246.899$ \\
				Gender: Male & $-27.590$ & $-145.227^{+}$ & $-143.429$ & $1337.644$ & $734.468$ & $4207.260$ \\
				Marital: Single & $-95.780^{*}$ & $-13.503$ & $404.998^{+}$ & $2045.943^{+}$ & $2717.375$ & $5780.085^{+}$ \\
				EoL ADL Score & $-124.500^{***}$ & $23.416$ & $155.252^{**}$ & $-966.058^{**}$ & $2018.021^{***}$ & $4261.245^{***}$ \\
				Home Owner: Yes & $-20.768$ & $203.895^{+}$ & $236.722$ & $-2347.149^{+}$ & $518.601$ & $2760.913$ \\
				Financial Distress & $-39.128^{***}$ & $-23.437^{**}$ & $-18.031^{***}$ & $-47.571$ & $6.440$ & $61.377$ \\
				Comorbidities & $42.204^{***}$ & $37.559^{+}$ & $63.469$ & $42.066$ & $392.149$ & $655.974$ \\
				Pandemic: COVID-19 & $244.313^{***}$ & $512.529^{***}$ & $876.805^{**}$ & $644.589$ & $6310.184^{**}$ & $12966.250^{**}$ \\
				\midrule
				Number of Observations & 2192 & 2192 & 2192 & 2192 & 2192 & 2192 \\
				$R^2$ & 0.043 & 0.016 & -0.833 & 0.031 & -0.340 & -1.570 \\
				\bottomrule
				\multicolumn{7}{l}{\footnotesize \textit{Note: $^{+} p<0.1$, $^{*} p<0.05$, $^{**} p<0.01$, $^{***} p<0.001$. Bootstrapped $xy$-pair standard errors.}}
			\end{tabular}%
		}
	\end{table*}
	
	\section{Robustness to the COVID-19 Pandemic Shock}
	\label{app:COVID}
	
	As established in our primary analysis, the structural architecture of EoL inequality was isolated by embedding a COVID-19 fixed effect into the models, effectively absorbing the aggregate macroeconomic friction of the 2020--2021 lockdowns. However, to deeply understand how this systemic stress test interacted with the individual drivers of welfare loss, we conducted a stratified temporal robustness check. We partitioned our synthesized empirical cohort into two distinct eras: the Pre-COVID-19 baseline (Wave 7 and 8, 2016--2019) and the Pandemic era (Wave 9, 2020--2021). The Marginal CATE on the net economic burden was independently re-estimated for each subset.
	
\begin{figure}[htbp]
		\centering
		\includegraphics[width=\columnwidth]{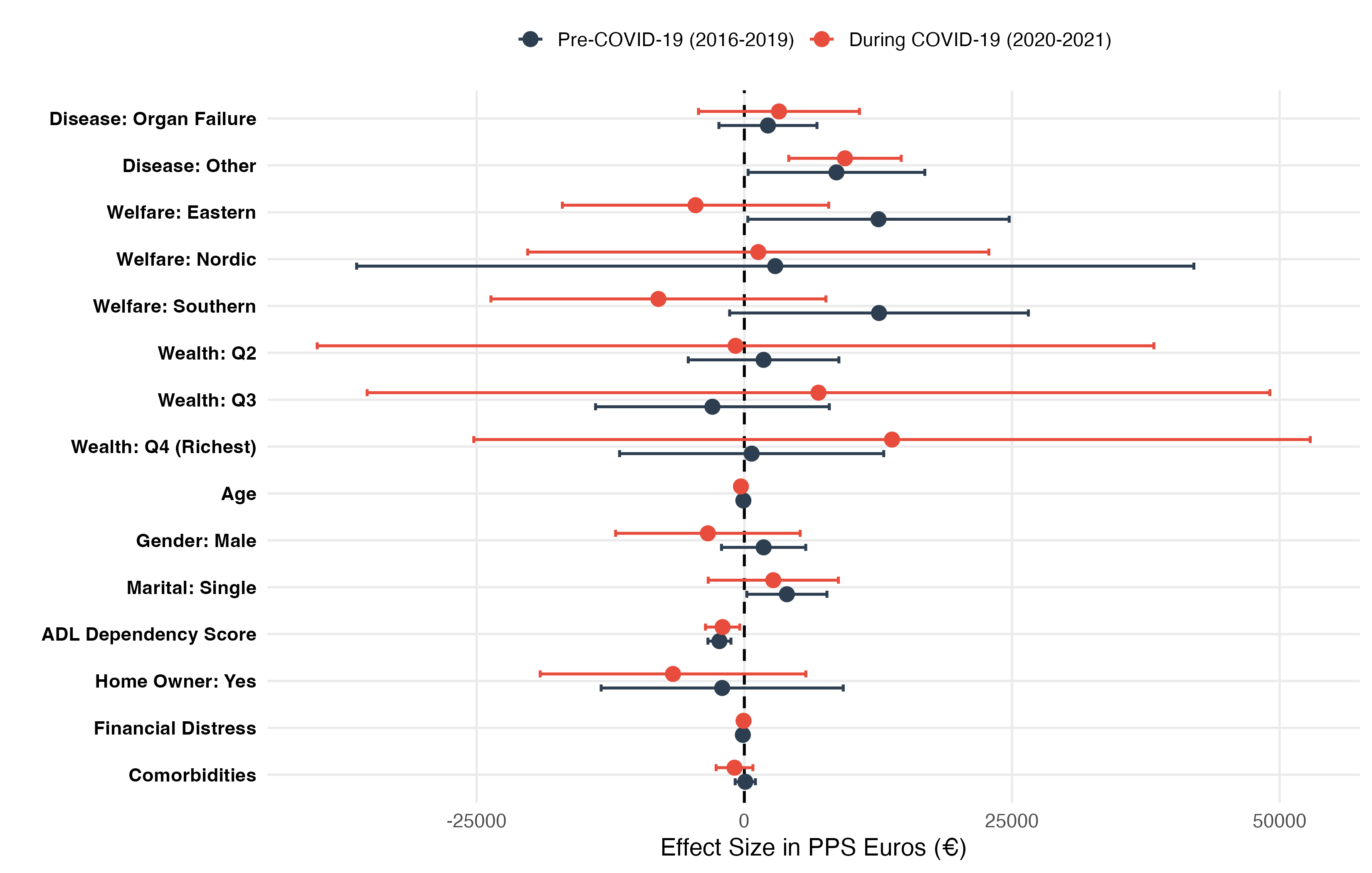}
		\caption{Temporal Robustness Analysis of Marginal CATE on Net Economic Burden (PPS €). Point estimates and 95\% CR2 confidence intervals are stratified by the Pre-COVID-19 (2016--2019) and During COVID-19 (2020--2021) periods.}
		\label{fig:covid_robustness}
	\end{figure}

	The comparative results, visualized in the forest plot (Figure \ref{fig:covid_robustness}), reveal a fascinating dichotomy between the clinical and socio-institutional drivers of the economic burden during an exogenous macro-shock. 
	
	Crucially, the underlying clinical and micro-socioeconomic determinants of the EoL burden—specifically age, the ADL dependency score, baseline comorbidities, and subjective financial distress—exhibit near-perfect alignment and overlapping confidence intervals between the pre-pandemic and pandemic cohorts. Furthermore, the massive economic penalty associated with non-cancer trajectories (`Disease: Other`) remained structurally entrenched and highly positive across both eras. This confirms that the mechanical cost of physical deterioration, as well as the systemic failure to accommodate prolonged dwindling trajectories, remained absolute and constant. This dynamic brilliantly validates the structural integrity of our baseline causal model, proving that disease-based inequality is a permanent architectural flaw rather than a transient pandemic artifact.
	
	Conversely, the pandemic introduced massive variance and structural displacement in the macro-institutional drivers. In the pre-COVID era, residing in Eastern and Southern European regimes drove severe economic penalties. During the COVID-19 period, the point estimates for these regional penalties converge toward the baseline, accompanied by an explosion in variance. This dynamic perfectly captures the reality of a systemic exogenous shock. During the lockdowns of 2020--2021, formal LTC services and nursing homes became largely inaccessible globally \citep{Werner2020, Grabowski2020}. 
	
	The explicit isolation of the COVID-19 shock yields an empirical demonstration of the fragility of LTC systems. From a microeconomic perspective, the pandemic induced a collision between perfectly inelastic demand and a collapsing formal supply. Severe EoL care represents an inflexible necessity; households cannot simply defer or opt out of providing assistance. The precise empirical mechanics of this collapse are documented by \citet{Bergmann2021}, who, utilizing SHARE data, measured the impact of the first pandemic wave on care networks. Their study demonstrates that containment measures generated a massive supply-side shock, causing the interruption or severe reduction of formal domiciliary care services for a significant proportion of the older European population. Faced with this sudden institutional vacuum, the data reveal that the deficit left by the system was entirely absorbed by informal family networks, driving an indiscriminate surge in the economic burden.
	
	Consequently, the ``institutional shield'' typical of the Continental and Nordic baselines temporarily collapsed, forcing all European households—regardless of their welfare regime—into a default familialistic model. The pandemic effectively acted as a great equalizer of burden, washing out the relative institutional penalties by uniformly degrading the formal care capacity of the entire continent \citep{Bambra2020}. Rather than indicating a structural failure of advanced welfare states, this empirical convergence reflects the unprecedented, mechanical reality of the global lockdowns. As formal care networks were physically or legally suspended to mitigate viral transmission, all European architectures predictably, and temporarily, regressed to a shared baseline of default familialism.
	
	Similarly, the drastic widening of confidence intervals for household wealth (Q2--Q4) during the COVID-19 period mathematically reflects the unprecedented price volatility and sheer unpredictability of navigating shadow care markets during global supply-chain disruptions. When formal networks collapse, even accumulated private wealth fails to reliably secure care. Ultimately, this robustness check confirms that our primary findings correctly capture the deeply rooted, ``peace-time'' architecture of EoL inequality across Europe, while the pandemic data accurately maps the chaotic market displacement of a systemic health crisis.
	
	\begin{table*}[htbp]
		\centering
		\caption{Baseline Characteristics and Observed Outcomes (Empirical Cohort)}
		\label{tab:empirical_cohort}
		\resizebox{0.88\textwidth}{!}{%
			\begin{tabular}{llcccc}
				\toprule
				& & \multicolumn{2}{c}{\textbf{Treated (N=1,864)}} & \multicolumn{2}{c}{\textbf{Untreated (N=328)}} \\
				\cmidrule(lr){3-4} \cmidrule(lr){5-6}
				\multicolumn{2}{l}{\textbf{Continuous Variables}} & \textbf{Mean} & \textbf{Std. Dev.} & \textbf{Mean} & \textbf{Std. Dev.} \\
				\midrule
				\multicolumn{2}{l}{Age} & 78.8 & 9.6 & 77.4 & 9.5 \\
				\multicolumn{2}{l}{Number of Children} & 2.2 & 1.5 & 1.7 & 1.3 \\
				\multicolumn{2}{l}{Financial Distress} & 8.7 & 24.5 & 7.0 & 20.5 \\
				\multicolumn{2}{l}{Number of Comorbidities} & 2.7 & 1.9 & 3.0 & 2.0 \\
				\multicolumn{2}{l}{ADL Dependency Score} & 3.3 & 2.5 & 2.9 & 2.6 \\
				\multicolumn{2}{l}{Shadow Hourly Wage (€)} & 18.4 & 9.1 & 14.8 & 8.5 \\
				\multicolumn{2}{l}{Observed OOP Costs (PPS)} & 2,404.1 & 6,559.5 & 1,840.1 & 4,426.0 \\
				\multicolumn{2}{l}{Observed Care Hours} & 2,111.7 & 2,173.7 & 2,006.8 & 2,164.4 \\
				\midrule
				\multicolumn{2}{l}{\textbf{Categorical Variables}} & \textbf{N} & \textbf{Pct. (\%)} & \textbf{N} & \textbf{Pct. (\%)} \\
				\midrule
				\textbf{Gender} & Female & 876 & 47.0 & 164 & 50.0 \\
				& Male & 988 & 53.0 & 164 & 50.0 \\
				\textbf{Marital Status} & Partnered & 1,068 & 57.3 & 171 & 52.1 \\
				& Single & 796 & 42.7 & 157 & 47.9 \\
				\textbf{Education (ISCED)} & 0 & 253 & 13.6 & 39 & 11.9 \\
				& 1 & 498 & 26.7 & 67 & 20.4 \\
				& 2 & 344 & 18.5 & 70 & 21.3 \\
				& 3 & 463 & 24.8 & 105 & 32.0 \\
				& 4 & 81 & 4.3 & 14 & 4.3 \\
				& 5 & 219 & 11.7 & 33 & 10.1 \\
				& 6 & 6 & 0.3 & 0 & 0.0 \\
				\textbf{Homeowner} & No & 1,001 & 53.7 & 219 & 66.8 \\
				& Yes & 863 & 46.3 & 109 & 33.2 \\
				\textbf{Wealth Quartile} & Q1 (Poorest) & 452 & 24.2 & 96 & 29.3 \\
				& Q2 & 432 & 23.2 & 116 & 35.4 \\
				& Q3 & 486 & 26.1 & 62 & 18.9 \\
				& Q4 (Richest) & 494 & 26.5 & 54 & 16.5 \\
				\textbf{Welfare Regime} & Continental & 330 & 17.7 & 43 & 13.1 \\
				& Eastern & 593 & 31.8 & 217 & 66.2 \\
				& Nordic & 268 & 14.4 & 24 & 7.3 \\
				& Southern & 673 & 36.1 & 44 & 13.4 \\
				\textbf{Cause of Death} & Cancer & 690 & 37.0 & 79 & 24.1 \\
				& Organ Failure & 632 & 33.9 & 149 & 45.4 \\
				& Other & 542 & 29.1 & 100 & 30.5 \\
				\bottomrule
			\end{tabular}%
		}
	\end{table*}
	
	\begin{table*}[htbp]
		\centering
		\caption{Potential Outcomes and Demographics (Full ATE Digital Twin Cohort)}
		\label{tab:ate_cohort}
		\resizebox{0.88\textwidth}{!}{%
			\begin{tabular}{llcccc}
				\toprule
				& & \multicolumn{2}{c}{\textbf{Standard Care ($Y_0$) (N=2,192)}} & \multicolumn{2}{c}{\textbf{Palliative Care ($Y_1$) (N=2,192)}} \\
				\cmidrule(lr){3-4} \cmidrule(lr){5-6}
				\multicolumn{2}{l}{\textbf{Continuous Variables}} & \textbf{Mean} & \textbf{Std. Dev.} & \textbf{Mean} & \textbf{Std. Dev.} \\
				\midrule
				\multicolumn{2}{l}{Age} & 78.6 & 9.6 & 78.6 & 9.6 \\
				\multicolumn{2}{l}{Number of Children} & 2.1 & 1.4 & 2.1 & 1.4 \\
				\multicolumn{2}{l}{Financial Distress} & 8.5 & 23.9 & 8.5 & 23.9 \\
				\multicolumn{2}{l}{Number of Comorbidities} & 2.7 & 1.9 & 2.7 & 1.9 \\
				\multicolumn{2}{l}{ADL Dependency Score} & 3.2 & 2.5 & 3.2 & 2.5 \\
				\multicolumn{2}{l}{Shadow Hourly Wage (€)} & 17.9 & 9.1 & 17.9 & 9.1 \\
				\multicolumn{2}{l}{Potential OOP Costs (PPS)} & 2,676.8 & 4,278.5 & 2,276.7 & 6,090.7 \\
				\multicolumn{2}{l}{Potential Care Hours} & 2,455.9 & 1,882.3 & 2,123.5 & 2,108.9 \\
				\midrule
				\multicolumn{2}{l}{\textbf{Categorical Variables}} & \textbf{N} & \textbf{Pct. (\%)} & \textbf{N} & \textbf{Pct. (\%)} \\
				\midrule
				\textbf{Gender} & Female & 1,040 & 47.4 & 1,040 & 47.4 \\
				& Male & 1,152 & 52.6 & 1,152 & 52.6 \\
				\textbf{Marital Status} & Partnered & 1,239 & 56.5 & 1,239 & 56.5 \\
				& Single & 953 & 43.5 & 953 & 43.5 \\
				\textbf{Education (ISCED)} & 0 & 292 & 13.3 & 292 & 13.3 \\
				& 1 & 565 & 25.8 & 565 & 25.8 \\
				& 2 & 414 & 18.9 & 414 & 18.9 \\
				& 3 & 568 & 25.9 & 568 & 25.9 \\
				& 4 & 95 & 4.3 & 95 & 4.3 \\
				& 5 & 252 & 11.5 & 252 & 11.5 \\
				& 6 & 6 & 0.3 & 6 & 0.3 \\
				\textbf{Homeowner} & No & 1,220 & 55.7 & 1,220 & 55.7 \\
				& Yes & 972 & 44.3 & 972 & 44.3 \\
				\textbf{Wealth Quartile} & Q1 (Poorest) & 548 & 25.0 & 548 & 25.0 \\
				& Q2 & 548 & 25.0 & 548 & 25.0 \\
				& Q3 & 548 & 25.0 & 548 & 25.0 \\
				& Q4 (Richest) & 548 & 25.0 & 548 & 25.0 \\
				\textbf{Welfare Regime} & Continental & 373 & 17.0 & 373 & 17.0 \\
				& Eastern & 810 & 37.0 & 810 & 37.0 \\
				& Nordic & 292 & 13.3 & 292 & 13.3 \\
				& Southern & 717 & 32.7 & 717 & 32.7 \\
				\textbf{Cause of Death} & Cancer & 769 & 35.1 & 769 & 35.1 \\
				& Organ Failure & 781 & 35.6 & 781 & 35.6 \\
				& Other & 642 & 29.3 & 642 & 29.3 \\
				\bottomrule
			\end{tabular}%
		}
	\end{table*}
	
\end{document}